\documentclass[review]{elsarticle}

\usepackage[margin=2.7cm]{geometry} % by courtesy of Mico
\usepackage{lineno,hyperref}
\usepackage{graphicx}
\usepackage{subcaption}
\graphicspath{ {./images/} }

\journal{Journal of \LaTeX\ Templates}

\usepackage{changes}
\usepackage{xcolor}
\usepackage{lipsum}
\usepackage{amsmath, nccmath}   % load the AMS math package
\usepackage{amsthm}    % allows AMS definitions for theorems, proofs, etc.
\usepackage{amssymb}   % allows the use of AMS symbols like blackboard bold
\usepackage{verbatim}

\usepackage{algorithmicx}
\usepackage{algorithm}
\usepackage[normalem]{ulem}
\usepackage{booktabs} % To thicken table lines
\usepackage{tabularx}
\DeclareMathOperator*{\argmin}{arg\,min}

\newcommand{\otoprule}{\midrule[\heavyrulewidth]}

\makeatletter
\newenvironment{breakablealgorithm}
  {% \begin{breakablealgorithm}
   \begin{center}
     \refstepcounter{algorithm}% New algorithm
     \hrule height.8pt depth0pt \kern2pt% \@fs@pre for \@fs@ruled
     \renewcommand{\caption}[2][\relax]{% Make a new \caption
       {\raggedright\textbf{\fname@algorithm~\thealgorithm} ##2\par}%
       \ifx\relax##1\relax % #1 is \relax
         \addcontentsline{loa}{algorithm}{\protect\numberline{\thealgorithm}##2}%
       \else % #1 is not \relax
         \addcontentsline{loa}{algorithm}{\protect\numberline{\thealgorithm}##1}%
       \fi
       \kern2pt\hrule\kern2pt
     }
  }{% \end{breakablealgorithm}
     \kern2pt\hrule\relax% \@fs@post for \@fs@ruled
   \end{center}
 }
 \makeatother

\usepackage{caption}
\usepackage[labelformat=simple]{subcaption}

\DeclareCaptionLabelFormat{subcaptionlabel}{\normalfont(\textbf{#2}\normalfont)}
\usepackage{graphicx}

\usepackage{environ}
\NewEnviron{myequation}{%
\begin{equation}
\scalebox{0.9}{$\BODY$}
\end{equation}
}

\begin{document}

\begin{frontmatter}

\title{Data-driven deep learning algorithms for time-varying infection rates of COVID-19 and mitigation measures}

%\title{Elsevier \LaTeX\ template\tnoteref{mytitlenote}}
%\tnotetext[mytitlenote]{Fully documented templates are available in the elsarticle package on \href{http://%www.ctan.org/tex-archive/macros/latex/contrib/elsarticle}{CTAN}.}


\author[mymainaddress]{K.D. Olumoyin\corref{mycorrespondingauthor}}
\cortext[mycorrespondingauthor]{Corresponding author}
\ead{kayode.olumoyin@mtsu.edu}

\author[mymainaddress]{A.Q.M. Khaliq}
%\ead{abdul.khaliq@mtsu.edu}

\author[mysecondaryaddress]{K.M. Furati}
%\ead{kmfurati@kfupm.edu.sa}

\address[mymainaddress]{Department of Mathematical Sciences,
Middle Tennessee State University,  Murfreesboro, TN 37132, USA}
\address[mysecondaryaddress]{Department of Mathematics,
 King Fahd University of Petroleum and Minerals, Dhahran 31261, Saudi Arabia}

\begin{abstract}
Epidemiological models with constant parameters may not capture satisfactory infection patterns in the presence of pharmaceutical and non-pharmaceutical mitigation measures during a pandemic, since infectiousness is a function of time. 
In this paper, an Epidemiology-Informed Neural Network algorithm is introduced to learn the time-varying transmission rate for the COVID-19 pandemic in the presence of various mitigation scenarios. 
There are asymptomatic infectives, mostly unreported, and the proposed algorithm learns the proportion of the total infective individuals that are asymptomatic infectives. Using cumulative and daily reported cases of the symptomatic infectives, we simulate the impact of non-pharmaceutical mitigation measures such as early detection of infectives, contact tracing, and social distancing on the basic reproduction number. We demonstrate the effectiveness of vaccination on the transmission of COVID-19.
The accuracy of the proposed algorithm is demonstrated using error metrics in the data-driven simulation for COVID-19 data of Italy, South Korea, the United Kingdom, and the United States.
\end{abstract}

\begin{keyword}
deep-learning \sep asymptotic population \sep COVID-19 \sep mitigation measures \sep time-varying transmission rate \sep reproduction number

%\texttt{elsarticle.cls}\sep \LaTeX\sep Elsevier \sep template
%\MSC[2010] 00-01\sep  99-00
\end{keyword}

\end{frontmatter}

%\linenumbers

%%%%%%%%%%%%%%%%%%%%%%%%%%%%%%%%%%%%%%%%%%%%%%%%%%%%%%%%%%%%%%%%%%%%%%%%%%%%%
\section{Introduction}
%%%%%%%%%%%%%%%%%%%%%%%%%%%%%%%%%%%%%%%%%%%%%%%%%%%%%%%%%%%%%%%%%%%%%%%%%%%%%

In December 2019, a new respiratory illness began to spread throughout Wuhan, China. The virus responsible for this illness is the SARS-CoV-2 and the disease is called COVID-19 %The paper should be referred to in numerical order. The document order has been corrected, please confirm
 \cite{WHO2021}. It quickly spread through Wuhan, a city of 11 million people in Hubei province. It infected tens of thousands of people over the ensuing weeks. China imposed major restrictions on travel and work, and by the end of February, cases of COVID-19 had slowed inside the country while spiking all over the world. COVID-19 data from different countries reflects various mitigation measures \cite{Lin2020,Tam2020}, such as lockdown, social distancing, early detection of infectives, contact tracing, and vaccination~\cite{Liu20202, Neves2020, Eikenberry2020}.
Many data-driven approaches in infectious disease modeling are linear models. When using linear regression, statistical methods such as Auto Regressive Moving Average (ARIMA) and Moving Average (MA) rely on assumptions which make it impossible to forecast transmission rate at any given time during a pandemic \cite{Chimula2020}. 
Time-varying transmission rates have been suggested to efficiently model the spread of COVID-19.
For example, fast methods for estimating time-varying transmission rate were introduced in~\cite{Jagan2020}; however, they reported that their method suffers from extreme sensitivity to noise.
In~\cite{Magri2020}, a first-principle machine learning approach was presented to predict time-dependent parameters, but these parameters require good initial guesses.
In March and April 2020, many countries instituted widespread lockdown \cite{Yale2020}.
A model-fitting approach for lockdown and lockdown relaxation is presented in \cite{Tepekule2021}, which requires good estimation of the model parameters as well as quantification of the impact of relaxation.
In \cite{Chowell2020}, the time-varying reproduction number $\mathcal{R}_t$ is estimated for counties in Georgia, USA, with a $95\%$ confidence credible interval.

The first epidemiology model, the SIR model, was presented by Kermack and McKendrick in 1927 \cite{Kermack1921}. 
The SIR model has inspired several epidemiological studies of diseases like, Malaria and Dengue fever \cite{Stolerman2015} and recently COVID-19.
A widely used threshold parameter for the spread or extinction of an infectious disease in an epidemiology model is the basic reproduction number \cite{Liu2019}.
It is defined as the average number of persons an infected person can infect.
When the basic reproduction number is less than one, the infectious disease vanishes.
In the SIR model \cite{Kermack1921}, the basic reproduction number is computed as the ratio of the transmission rate to the recovery rate.
In this paper, we adopt a variant of the asymptomatic-SIR model presented in \cite{Gaeta2021}.
When the transmission and recovery rates are constants, the basic reproduction number is given by the ratio of the transmission rate to a weighted sum of the symptomatic and asymptomatic recovery rates.
However, When the transmission rate is time-varying, we use a modified reproduction,
which we call the time-varying reproduction $\mathcal{R}_t$.
This time-varying reproduction number, $\mathcal{R}_t$, demonstrates the spread pattern of COVID-19 throughout the duration of the pandemic.

There is an asymptomatic period for every infective individual in the range of 7 to 14 days \cite{Liu2020}.
There are also asymptomatic infectives that never show symptoms but are infectious \cite{Gaeta2021}.
Early studies of the spread of COVID-19 shows that some of the infectives are asymptomatic infectives \cite{He2020, CDC2020} and they are mostly unreported in the publicly available data \cite{Gaeta2021}. 
In \cite{LongQ2020}, it was reported that the asymptomatic infectives can spread the virus efficiently, and they are the silent spreaders of COVID-19, which has caused difficulties in the control of the pandemic. Early in the pandemic, the Centers for Disease Control and Prevention (CDC) estimates the proportion of the asymptomatic infectives to be $40\%$ of the total infectives in the USA \cite{CDC2020}.
A high population proportion of asymptomatic infectives was estimated in \cite{He2020} for China and Singapore. In \cite{LongQ2020}, the proportion of Asymptomatic infectious patients in Wanzhou district before 10 April 2020 was $20\%$.  \cite{Gaeta2021} reported $10\%$ of the total infectives were asymptomatic in northern Italy. In a study conducted in England from June through September 2020 and in Spain from 27 April to 11 May 2020, the proportions of asymptomatic infectives in England and Spain were reported to be $32.4\%$ and $33.0\%$ respectively \cite{Oran2021}.

Deep learning \cite{LeCun2015} and Neural networks have found applications in function approximation tasks, since neural networks are known to be universal approximators of continuous functions \cite{Cybenko1989, Hornik1991}.
Feedforward neural networks (FNN) have been used to learn approximate solutions of differential equations.
In~\cite{Wenhua2020}, FNN was combined with the traditional Cox model for survival analysis to predict the clinical outcome of COVID-19 patients.
In~\cite{Raissi2019}, FNN was used to develop differential equation solvers and parameter estimators by constraining the residual.
This FNN is called the Physics Informed Neural Network (PINN).
PINN has been used to simulate pandemic spread, see \cite{RaisiM2019}, where the model parameters were taken to be constants \cite{Raissi2019, Raissi2020}, PINN was used to solve nonlinear partial differential equations from data. PINN has been used to solve system of ordinary differential equations~\cite{Yazdani2020} and system of fractional differential equations \cite{Kharazmi2021}.
In \cite{Long2020}, an algorithm that combines PINN together with LSTM is presented to solve an epidemiological model and identify weekly and daily time-varying parameters.

To overcome the limitations of statistical approaches, we present an Epidemiology-Informed Neural Network (EINN) inspired by applying a PINN to epidemiology models.
Given that it may not be possible to know the most accurate form of a time-varying transmission rate, EINN algorithms is a viable option to learn time-varying transmission rate and to detect the impact of mitigation measures from data.
The EINN loss function is extended to include some known epidemiology facts about infectious diseases.
To detect hidden details in the training data, a cubic spline interpolation is used to generate sufficient training data.
The proposed EINN algorithm can capture the dynamics of the spread of the disease and the influence of various mitigation measure.
Since asymptomatic infectives population is unreported in the publicly available data \cite{Dong2020}.
EINN algorithm learns asymptomatic infectives population by training on symptomatic infectives data that are available in the reported public data.

 The paper is organized as follows. 
 In Section \ref{A-SIR}, we introduce and discuss the asymptomatic-SIR model, the neural network structure of EINN and the EINN algorithm for time-varying transmission rate. 
In Section \ref{results}, data-driven simulation results for constant transmission rates, data-driven simulation results for pharmaceutical and non-pharmaceutical mitigation measures, and data-driven simulation results for time-varying transmission rates are presented. 
In Section \ref{discuss}, we discuss the mitigation measures, vaccination efficacy, the time-varying transmission results and error metrics for data-driven simulation.
Finally, a summary of the results in this paper is presented in Section \ref{conclusion}.

%An asymptomatic-SIR model is a modification to the traditional SIR infectious disease model \cite{Kermack1921}, that includes asymptomatic infectives and asymptomatic recovered compartments \cite{Gaeta2021}.
%We introduce modifications to the asymptomatic-SIR model by including a vaccination compartment determined by a vaccination rate, time-varying transmission rate, and a probability term that determines the portion of the total infectives that are asymptomatic infectives.

%%%%%%%%%%%%%%%%%%%%%%%%%%%%%%%%%%%%%%%%%%%%%%%%%%%%%%%%%%%%%%%%%%%%%%%%%%%%%
\section{Materials and Methods}
\label{A-SIR}
%%%%%%%%%%%%%%%%%%%%%%%%%%%%%%%%%%%%%%%%%%%%%%%%%%%%%%%%%%%%%%%%%%%%%%%%%%%%%
\subsection{Asymptomatic-SIR Model} 
%%%%%%%%%%%%%%%%%%%%%%%%%%%%%%%%%%%%%%%%%%%%%%%%%%%%%%%%%%%%%%%%%%%%%%%%%%%%%

The asymptomatic-SIR model introduced in \cite{Gaeta2021} assumes that some of the infectives are asymptomatic infectives.
This group is infectious despite not showing symptoms of COVID-19, probably are not tested, and are usually unreported in the various publicly available data.
%In Figure~\ref{fig1}, the interactions between the different compartments of the asymptomatic-SIR model are shown.

\begin{figure}[H]
	\centering
	\includegraphics[width=12 cm]{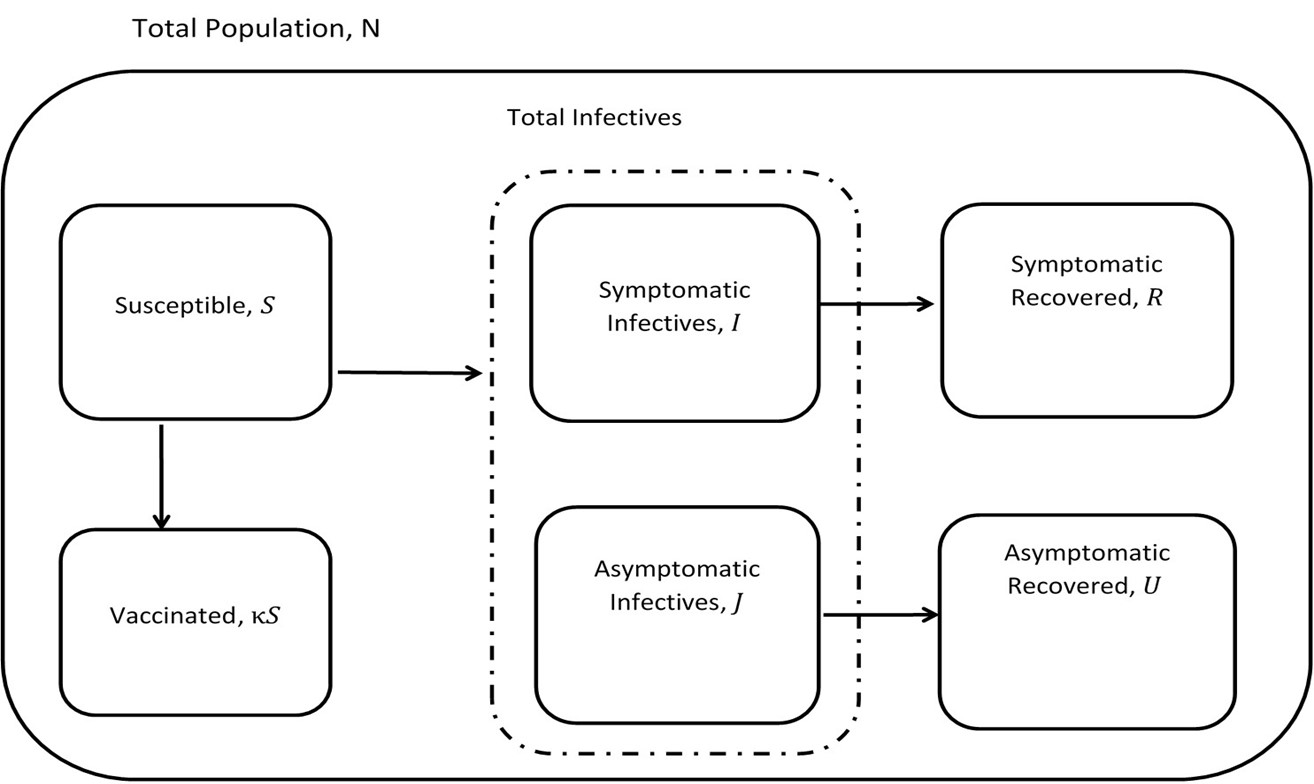}
	\caption{Compartments in Asymptomatic-SIR model with vaccination} \label{fig1}
\end{figure}

The asymptomatic-SIR model considers the following population compartments: the Susceptible $(S)$, the symptomatic Infectives $(I)$ which correspond to the reported infectives in the publicly available data, and the asymptomatic Infectives $(J)$ which correspond to the unreported infectives.
The total infectives are $I + J$.
The rest of the compartments are the symptomatic Recovered $(R)$ and the asymptomatic Recovered $(U)$. 
The symptomatic Infectives $(I)$ recover at the rate $\gamma$, and the asymptomatic Infectives $(J)$ recover at the rate  $\mu$.
$I$ recover through isolation in the hospital or at home. On the other hand, the $J$ recover spontaneously.
%In \cite{Gaeta2021}, it was reported that typically, $\mu < \gamma$. so that the typical recovery time $\mu^{-1}>\gamma^{-1}$.
The vaccinated population, $(V = \kappa S)$, is a loss from the susceptible compartment: they are added to the recovered compartments.
$\beta(t)$  is the time-varying transmission rate, it usually depends on the infection vector.
In the COVID-19 pandemic, $\beta(t)$ depends also on contacts between individuals. 
$\kappa$ is the average percentage of individuals that are vaccinated daily.
$\xi$ represents the probability that an infective individual is reported, while $(1 - \xi)$ is the probability that an infective is an asymptomatic infective.
The portion of the total infectives that are symptomatic and reported corresponds to $\xi (I+J)$. 
On the other hand, $(1-\xi)(I+J)$ represents the asymptomatic infectives.  
$N$ represents the total population \eqref{continuity_eqn}.
It is assumed that $N$ does not change throughout the pandemic and that infective individuals are immediately infectious.
The dynamics of the interactions between the compartments in Figure~\ref{fig1} can be represented by the following system of ordinary differential equations with time-varying transmission rate $\beta(t)$. 
%\eqref{Asymptomatic-SIRD_0}.

\begin{equation}\label{Asymptomatic-SIRD_0} 
	\begin{split}
		\frac{d S(t)}{d t}   & =  -\frac{1}{N}\beta(t) \Big(I(t) + J(t)\Big) S(t)  - \kappa S(t)\\
		\frac{d I(t)}{d t}   & =  \frac{1}{N}\beta(t) \xi \Big(I(t) + J(t)\Big) S(t) - \gamma I(t) \\
		\frac{d J(t)}{d t}   & =  \frac{1}{N}\beta(t) \Big(1 - \xi\Big) \Big(I(t) + J(t)\Big) S(t) - \mu J(t) \\
		\frac{d R(t)}{d t}   & =  \gamma I(t) + \kappa \xi S(t) \\
		\frac{d U(t)}{d t}   & =  \mu  J(t) + \kappa (1 - \xi) S(t). 
	\end{split}
\end{equation}

The continuity equation is given by

\begin{equation}\label{continuity_eqn}
	\begin{split}
		N(t) = S(t) + I(t) + J(t) + R(t) + U(t), \hspace{5 MM} t\geq t_0.
	\end{split}
\end{equation}

 The initial conditions are  denoted by $S(t_0) = S_0 $, $I(t_0) = I_0$, $J(t_0) = J_0$, $R(t_0) = R_0$, and $U(t_0) = U_0$, where $t\geq t_0$ represent time in days and $t_0$ is the start date of the pandemic in the model. The model parameters are summarized in Table~\ref{Table:0}.

\begin{table}[H]
\footnotesize
\begin{tabular}{lcccc}
\addlinespace
\toprule
{\bf Parameter} & {\bf Notation} & {\bf Range} & {\bf Remark} & {\bf Reference}\\
\otoprule
{Baseline transmission rate} & {$\beta_0$} & {[0,1)} & {fitted using early data} & {\citep*{Liu2020, Magri2020}}  \\
{Probability that an Infected person is reported} & {$\xi$} & {$[0,1)$} & {constant}  & {\cite{Gaeta2021}}   \\
{Proportions of daily vaccinated individuals} & {$\kappa$} & {$[0,1)$} & {constant}  & {\cite{Liu2020, Liu20202}}   \\
{recovery rate of symptomatic infectives } & ${\gamma}$ & {[0,1)} & {constant}  & {\cite{Gaeta2021}}   \\
{recovery rate of asymptomatic infectives} & ${\mu}$ & {[0,1)} & {constant}  & {\cite{Gaeta2021}}   \\ 
\midrule
\end{tabular}
\caption{Summary table of parameters in~\eqref{Asymptomatic-SIRD_0}}
\label{Table:0}
\end{table}

%\begin{specialtable}[H]\scriptsize
%\tablesize{\small}
%\caption{Summary table of parameters in %Table 1 is not mentioned in this paper, please add it
% \eqref{Asymptomatic-SIRD_0}}
%\label{Table:0}
%\setlength{\cellWidtha}{\columnwidth/5-2\tabcolsep+0.6in}
%\setlength{\cellWidthb}{\columnwidth/5-2\tabcolsep-0.3in}
%\setlength{\cellWidthc}{\columnwidth/5-2\tabcolsep-0.5in}
%\setlength{\cellWidthd}{\columnwidth/5-2\tabcolsep+0.5in}
%\setlength{\cellWidthe}{\columnwidth/5-2\tabcolsep-0.3in}
%\scalebox{1}[1]{\begin{tabularx}{\columnwidth}{>{\PreserveBackslash\raggedright}m{\cellWidtha}>{\PreserveBackslash\raggedright}m{\cellWidthb}>{\PreserveBackslash\raggedright}m{\cellWidthc}>{\PreserveBackslash\raggedright}m{\cellWidthd}>{\PreserveBackslash\raggedright}m{\cellWidthe}}
%\toprule
%         \textbf{Parameter} & \textbf{Notation} & \textbf{Range} & \textbf{Remark} & \textbf{Reference} \\ 
%         \midrule
%         {Baseline transmission rate} & {$\beta_0$} & {[0,1)} & {fitted using early data} & {\cite{Liu2020, Magri2020}}  \\
%         {Probability that an Infected person is reported} & {$\xi$} & {$[0,1)$} & {constant}  & {\cite{Gaeta2021}}   \\
%         {Proportions of individuals that are vaccinated daily} & {$\kappa$} & {$[0,1)$} & {constant}  & {\cite{Liu2020, Liu20202}}   \\
%         {recovery rate of symptomatic infectives } & ${\gamma}$ & {[0,1)} & {constant}  & {\cite{Gaeta2021}}   \\
%         {recovery rate of asymptomatic infectives} & ${\mu}$ & {[0,1)} & {constant}  & {\cite{Gaeta2021}}   \\ \bottomrule
%\end{tabularx}}
%\end{specialtable}

\subsection{Time-Varying Transmission Rate}
Time-varying transmission rate $\beta(t)$ in \eqref{Asymptomatic-SIRD_0} incorporates the impact of public health actions and the public response to the actions \cite{HeD2013, Lin2020}. 
The formulation of $\beta(t)$ in \cite{HeD2013} includes temperature. This parameter is not considered in the formulation presented in~\cite{Lin2020}, since there is no evidence that temperature plays a role in the transmission of COVID-19. 
Early in the transmission of COVID-19, the major public health action was lockdown, which was followed by other measures such as social distancing, contact tracing, masking, early detection of infectives and so on. We chose a formulation of $\beta(t)$ that strongly reflects the pre and post-lockdown periods.
In \cite{Tepekule2021} a sigmoid function is used to model a time-dependent decrease in the transmission of COVID-19. 
In \cite{Gaeta2021}, a piecewise constant function is used to model $\beta(t)$. A piecewise time-varying transmission rate \eqref{gaeta_eqn} is used to learn a time-dependent transmission rate $\beta$ in eq. \eqref{Asymptomatic-SIRD_0}. In~\cite{Gaeta2021}, the piecewise $\beta(t)$  is defined as follows, 

\begin{equation}\label{gaeta_eqn}
\beta(t) = \begin{cases}
\beta_0q_1 &  t \leq M_1\\
\beta_0q_2 & M_1 < t \leq M_2\\
\beta_0q_3 & M_2 < t \leq M_3\\
\beta_0q_4 & M_3 < t \leq M_4\\
& \vdots\\
\beta_0q_n & M_{n} < t. 
\end{cases}
\end{equation}

The goal of the parameters $q_1, \ldots, q_n$ in~\eqref{gaeta_eqn} is to capture the exponential decrease observed in the transmission rate $\beta(t)$. We choose $M_1, \ldots, M_n$ in order to partition the pandemic timeline, according to the onset of various mitigation measures.

We also formulate $\beta(t)$ following the approach presented in \cite{Liu2020, olumoyin2021}. An exponentially decreasing function is used to represent the transmission rate  $\beta(t)$ in \eqref{Asymptomatic-SIRD_0} to model the impact of lockdown. 
\begin{equation}\label{webb_eqn}
\beta(t) = \begin{cases}
\beta_0, & 0 \leq t \leq K,\\
\beta_0 \exp{(-\eta(t-K))}, & K < t
\end{cases}
\end{equation}
\noindent where $K$ signifies the onset of government intervention including isolation, quarantine and lockdown. $\eta$ is the rate at which human contact decreases. 
We denote $K$ to be the number of days between the date of the first reported case of COVID-19 and the date lockdown was instituted.

%\begin{specialtable}[H]\scriptsize
%\tablesize{\small}
%\caption{\hl{Summary table of parameters in} %Table 1 is not mentioned in this paper, please add it
% \eqref{Asymptomatic-SIRD_0}}
%\label{Table:0}
%\setlength{\cellWidtha}{\columnwidth/5-2\tabcolsep+0.6in}
%\setlength{\cellWidthb}{\columnwidth/5-2\tabcolsep-0.3in}
%\setlength{\cellWidthc}{\columnwidth/5-2\tabcolsep-0.5in}
%\setlength{\cellWidthd}{\columnwidth/5-2\tabcolsep+0.5in}
%\setlength{\cellWidthe}{\columnwidth/5-2\tabcolsep-0.3in}
%\scalebox{1}[1]{\begin{tabularx}{\columnwidth}{>{\PreserveBackslash\raggedright}m{\cellWidtha}>{\PreserveBackslash\raggedright}m{\cellWidthb}>{\PreserveBackslash\raggedright}m{\cellWidthc}>{\PreserveBackslash\raggedright}m{\cellWidthd}>{\PreserveBackslash\raggedright}m{\cellWidthe}}
%\toprule
%         \textbf{Parameter} & \textbf{Notation} & \textbf{Range} & \textbf{Remark} & \textbf{Reference} \\ 
%         \midrule
%         {Baseline transmission rate} & {$\beta_0$} & {[0,1)} & {fitted using early data} & {\cite{Liu2020, Magri2020}}  \\
%         {Probability that an Infected person is reported} & {$\xi$} & {$[0,1)$} & {constant}  & {\cite{Gaeta2021}}   \\
%         {Proportions of individuals that are vaccinated daily} & {$\kappa$} & {$[0,1)$} & {constant}  & {\cite{Liu2020, Liu20202}}   \\
%         {recovery rate of symptomatic infectives } & ${\gamma}$ & {[0,1)} & {constant}  & {\cite{Gaeta2021}}   \\
%         {recovery rate of asymptomatic infectives} & ${\mu}$ & {[0,1)} & {constant}  & {\cite{Gaeta2021}}   \\ \bottomrule
%\end{tabularx}}
%\end{specialtable}

When the transmission rate in \eqref{Asymptomatic-SIRD_0} is assumed to be constant, ($\beta(t)=\beta)$, the basic reproduction number can be given by the ratio of the transmission rate to a weighted sum of the symptomatic and asymptomatic recovery rates. However, we observed that this under-estimate the basic reproduction number ($\mathcal{R}_0$) for the asymptomatic-SIR model Equation~\eqref{Asymptomatic-SIRD_0}. Assuming a disease-free equilibrium of~\eqref{Asymptomatic-SIRD_0}, given by 
\[ (S^{\ast}, I^{\ast}, J^{\ast}, R^{\ast}, U^{\ast}) = (S_0, 0, 0, 0, 0)\]
Applying the next generation matrix approach~\citep*{Driessche2002}, the basic reproduction number ($\mathcal{R}_0$) is obtained as the spectral radius of the next generation matrix $FV^{-1}$, where

\[
F = \begin{pmatrix}
\beta \xi & \beta \xi\\
\beta (1-\xi) & \beta (1-\xi)
\end{pmatrix},
\hspace{5 MM}
V = \begin{pmatrix}
\gamma & 0\\
0 & \mu
\end{pmatrix}.
\]

so that
\begin{equation}
\mathcal{R}_0 =  \frac{\beta \left( \xi \mu + (1-\xi)\gamma \right)}{\mu \gamma } \quad \xi \in (0,1).
\end{equation}

If $\xi = 0$, $\mathcal{R}_0 = \beta/\mu$, when all the infective population are asymptomatic.

If $\xi = 1$, $\mathcal{R}_0 = \beta/\gamma$, when all the infective population are symptomatic.

Using data from Italy, South Korea, and the United States starting from the date of the first reported cases in the respective countries to the day before vaccination data were reported.
The cumulative infective and recovered population data are observed to be non-exponential whenever a mitigation measure such as a comprehensive lockdown is detected in the data.
We take the total population $N$ to be $60.36 \times 10^{6}$, $51.64 \times 10^{6}$, and $328.2 \times 10^{6}$ in Italy, South Korea and the USA, respectively. In Figures \ref{italy_learn_params}a--\ref{usa_learn_params}a, $M_{\kappa}$ is zero and so $\kappa = 0$ for all the period from the first reported cases to the day before vaccination data are reported.
In addition to learning the parameters, EINN learns $\xi$, the probability that an infective is reported.
A high value of $\xi$ indicates a large number of reported infectives.

When the transmission rate is time-varying, we use a modified reproduction, which we call the time-varying reproduction $\mathcal{R}_t$. This time-varying reproduction number, $\mathcal{R}_t$, demonstrates the spread pattern of COVID-19 throughout the duration of the pandemic~\citep*{Gaeta2021}.

\begin{equation}\label{brn2}
\mathcal{R}_t =  \frac{\beta(t) \left( \xi \mu + (1-\xi)\gamma \right)}{\mu \gamma } \quad \xi \in (0,1).
\end{equation}

%We use the time-varying reproduction rate $\mathcal{R}_t$ as presented in \cite{Gaeta2021}  for the asymptomatic-SIR model \eqref{Asymptomatic-SIRD_0} given by,
%
%\begin{equation}\label{brn2}
%\mathcal{R}_t = \frac{\gamma \beta(t)}{\left[\gamma \left(\frac{I(t)}{I(t) + J(t)}\right) + \mu \left(1 - \frac{I(t)}{I(t) + J(t)}\right) \right]^2}.
%\end{equation}

%%%%%%%%%%%%%%%%%%%%%%%%%%%%%%%%%%%%%%%%%%%%%%%%%%%%%%%%%%%%%%%%%%%%%%%%%%%%%
\subsection{Neural Network Structure}
\label{EINN}

%%%%%%%%%%%%%%%%%%%%%%%%%%%%%%%%%%%%%%%%%%%%%%%%%%%%%%%%%%%%%%%%%%%%%%%%%%%%%
\subsubsection{Feedforward Neural Network (FNN)}

An FNN can be represented as a function of $L$ layers, $t$ input vector and an output $\mathcal{N}$
\begin{equation}
\mathcal{N}(t;\theta) = \sigma(W_{L}\sigma(\ldots \sigma(W_{2} \sigma(W_{1}t + b_1)+b_2)\ldots)+b_L),
\end{equation}
\noindent where $\theta$: = $(W_1, \ldots, W_L,b_1, $\ldots$, b_L)$. $W_k$, $k = 1, \ldots, L$, is the set of the neural network weight matrices while $b_k$, $k = 1, \ldots, L$, is the set of the bias vectors. $\sigma$ is the activation function. Given a collection of sample pairs $(t_j, u_j)$, $j = 1,\dots M$, where $u$ is some target function.
The goal is to find $\theta^{*}$ by solving the optimization problem
\begin{equation}\label{orig_lossFunc}
\theta^{*} = \argmin\limits_{\theta}\frac{1}{M}\sum_{j=1}^{M}||\mathcal{N}(t_j;\theta) - u_j||_2^2.
\end{equation}

The function $\frac{1}{M}\sum_{j=1}^{M}||\mathcal{N}(t_j;\theta) - u_j||_2^2$ on the right-hand side of \eqref{orig_lossFunc} is called the mean squared error (MSE) loss function.
A major task in training a network is to determine the suitable number of layers and the number of neurons per layer needed, the choice of activation function, and an appropriate optimizer for the loss function \cite{Goodfellow2016}.

%%%%%%%%%%%%%%%%%%%%%%%%%%%%%%%%%%%%%%%%%%%%%%%%%%%%%%%%%%%%%%%%%%%%%%%%%%%%%
\subsubsection{Epidemiology-Informed Neural Network (EINN)}

EINN is a type of Feedforward Neural Network that includes the known epidemiology dynamics in its loss function.
In this paper, EINN is adapted for the asymptomatic-SIR model \eqref{Asymptomatic-SIRD_0}, where the Mean Square Error (MSE) of this neural network's loss function includes the known epidemiology dynamics such as a lockdown, while other mitigation measures such as social distancing, and contact tracing are detected by the time-varying transmission rate.
The output of EINN are the learned solutions to the asymptomatic-SIR model \eqref{Asymptomatic-SIRD_0} denoted by $S(t_j; \theta; \lambda)$, $I(t_j; \theta; \lambda)$, $J(t_j; \theta; \lambda)$, $R(t_j; \theta; \lambda)$, $U(t_j; \theta; \lambda)$, $j=1,\ldots,M$. 
Where $\theta$ represent the neural network weights and biases and $\lambda$ represent the epidemiology parameters. $M$ is the number of training set.
The network representing the time-varying transmission rate is denoted by
$\beta(t_j;\phi;\eta)$, $j=1,\ldots,M$, The parameter $\phi$ represents the weights and biases of this network and $\eta$ is the exponential decay parameter.
The training data are generated using cubic spline and denoted by $\tilde{I}(t_j)$, $\tilde{R}(t_j)$, $j=1,\ldots,M$ and $\tilde{V}(t_j)$, $j=1,\ldots,M_{\kappa}$ from the given dataset. 
Here $M_{\kappa}$ is the number of vaccination days.
We observe that training data are not available for all the compartments in the asymptomatic-SIR model; however, EINN is able to capture the epidemiology interactions between the compartments because the epidemiology model residual is included in the MSE loss function.
The MSE loss function for EINN with the time-varying transmission rate is given by
\begin{fleqn}[\parindent]\label{loss_func}
\begin{equation}
\begin{split}
MSE &= \frac{1}{M}\sum_{j=1}^{M}||I(t_j; \theta; \lambda) - \tilde{I}(t_j)||_2^2 + \frac{1}{M}\sum_{j=1}^{M}||R(t_j; \theta; \lambda) - \tilde{R}(t_j)||_2^2 \\
& +  \frac{1}{M_{\beta}}\sum_{j=1}^{M_{\beta}}||\beta(t_j; \phi;\eta) - \tilde{\beta}(t_j)||_2^2\\
& +  \frac{1}{M_{\kappa}}\sum_{j=1}^{M_{\kappa}}||\kappa S(t_j;\theta; \lambda) - \tilde{V}(t_j)||_2^2\\
& + ||J(0;\theta;\lambda) - \tilde{J}(0)||_2^2 + ||U(0;\theta; \lambda) - \tilde{U}(0)||_2^2\\
& + \frac{1}{M}\sum_{i=1}^6\sum_{j=1}^{M}||L_i(t_j;\theta;\phi; \lambda;\eta)||_2^2,
\end{split}
\end{equation}
\end{fleqn}
where the residual $L_i$, $i=1,\ldots 6$ is as follows
\begin{fleqn}[\parindent]\label{resid_loss}
\begin{myequation}
\begin{split}
L_1(t_j;\theta;\phi;\lambda;\eta) &=\frac{d S(t_j; \theta;\lambda)}{d t_j} + \frac{1}{N}\beta(t_j; \phi;\eta) \Big(I(t_j; \theta;\lambda) + J(t_j; \theta;\lambda)\Big)S(t_j; \theta;\lambda) \\
&+ \kappa S(t_j; \theta;\lambda)\\
L_2(t_j;\theta;\phi;\lambda;\eta) &= \frac{d I(t_j; \theta;\lambda)}{d t_j} - \frac{1}{N}\beta(t_j; \phi;\eta) \xi \Big(I(t_j; \theta;\lambda) + J(t_j; \theta;\lambda)\Big)S(t_j; \theta;\lambda) \\
&+ \gamma I(t_j; \theta;\lambda)\\ 
L_3(t_j;\theta;\phi;\lambda;\eta) &= \frac{d J(t_j; \theta;\lambda)}{d t_j} - \frac{1}{N}\beta(t_j; \phi;\eta) \Big(1 - \xi \Big)\Big(I(t_j; \theta;\lambda) + J(t_j; \theta;\lambda)\Big)S(t_j; \theta;\lambda) \\
&+ \mu J(t_j; \theta;\lambda)\\ 
L_4(t_j;\theta;\phi;\lambda;\eta) &= \frac{d R(t_j; \theta;\lambda)}{d t_j}   -   \gamma I(t_j; \theta;\lambda) - \kappa \xi S(t_j; \theta;\lambda) \\
L_5(t_j;\theta;\phi;\lambda;\eta) &=\frac{d U(t_j; \theta;\lambda)}{d t_j}   -  \mu  J(t_j; \theta;\lambda) - \kappa(1 - \xi)S(t_j; \theta;\lambda) \\
L_6(t_j;\theta;\phi;\lambda;\eta) &=  N - (S(t_j; \theta;\lambda) + I(t_j; \theta;\lambda) + J(t_j; \theta;\lambda) + R(t_j; \theta;\lambda) + U(t_j; \theta;\lambda)).
\end{split}
\end{myequation}
\end{fleqn}

\begin{figure}[H]
\centering
\includegraphics[width=14 cm]{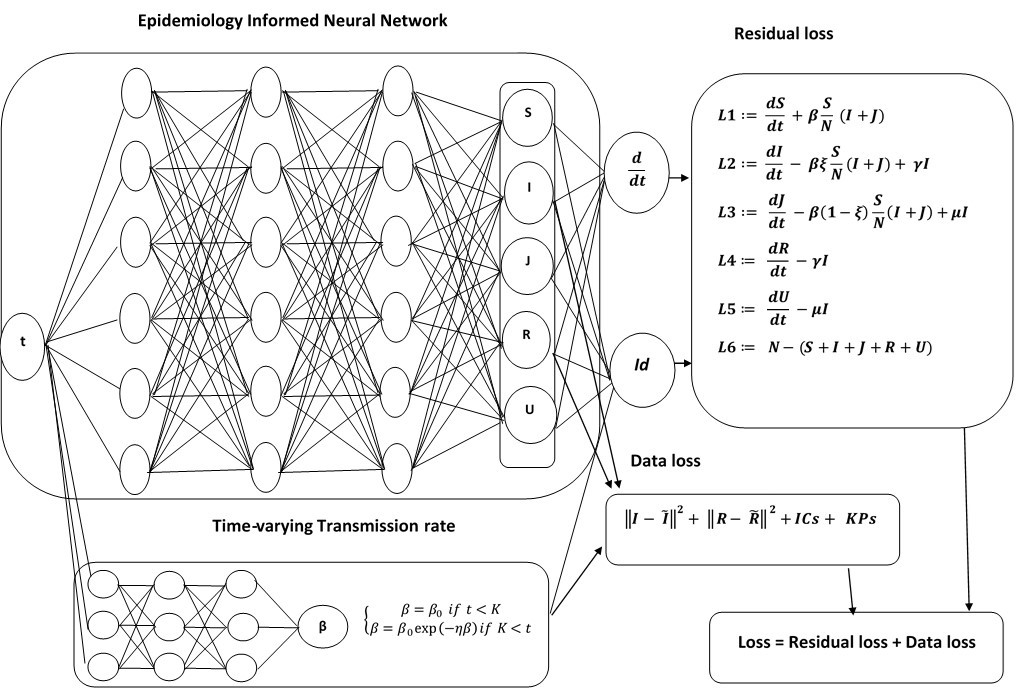}
\caption{Schematic diagram of the Epidemiology-Informed Neural Network with nonlinear time-varying transmission rate. The term $KPs$ represent the known dynamics in the transmission rates pattern and $ICs$ represent the initial condition for the asymptomatic population. \label{Sch_A_SIR}}
\end{figure}

In Figure~\ref{Sch_A_SIR}, EINN includes the time-varying infection as an output of the neural network.
$ICs$ represents the loss in the neural network output for the asymptomatic infectives $J(0;\theta)$ and the asymptomatic recovered $U(0;\theta)$ at $t=0$. 
%actual asymptomatic infectives $J(0)$ determined according to the formula $J(0) = (1-\xi) I(0)/\xi$.
%And the loss at $t=0$ in the neural network output for the asymptomatic recovered $U(0;\theta)$ and the actual asymptomatic recovered $U(0)$ determined according to the formula $U(0) = (1-\xi) R(0)/\xi$.
$KPs$ represent the known dynamics in the transmission rates pattern.
$M$ is the number of training points. M does not necessarily correspond to the number of available data. $M$ is generated by fitting the data with cubic splines.
For instance, $\tilde{I}(t_j)$, $j = 1, \ldots, M$ is the training data for the infectives after fitting with an interpolation function.
$M_{\beta}$ is the number of training points used to enforce the known dynamics of the transmission rates pattern.
%On the other hand, $M_{\kappa}$ is the number of vaccination days in the model, 
Since $\kappa$ is the average percentage of individuals that are vaccinated daily, $M_{\kappa}$ is the number of days $\kappa$ is not zero.
$\tilde{V}(t_j) = \kappa \tilde{S}(t_j)$, $j=1,\ldots,M_{\kappa}$, is the daily vaccination data.
% for days $\kappa$ is not zero.
The input to EINN is $t_j$, $j=1,\ldots,M$.
%The epidemiological model is embedded in the residual given in. \eqref{resid_loss}.
%EINN solutions are not unique. 
To achieve good accuracy in the neural network, we tune the hyperparameters; such as the number of layers, number of training points, and the learning rate.
In all the simulations presented in this paper, we used 4 hidden layers, 64 neurons per layer, and the training loss was minimized in 40,000 iterations.
Cubic splines are used to generate 3000 training points from the original dataset.
The loss function is minimized by a gradient-based optimizer such as the adam optimizer \cite{Kingma2017}.

%%%%%%%%%%%%%%%%%%%%%%%%%%%%%%%%%%%%%%%%%%%%%%%%%%%%%%%%%%%%%%%%%%%%%%%%%%%%%

\section{Results}
\label{results}

\subsection{Data-Driven Simulation Results for Constant Transmission Rates} 

\begin{figure}[H]
	\centering
	\begin{subfigure}[b]{\textwidth}
		\captionsetup{width=0.5\textwidth}
		\centering
		\includegraphics[width=17 cm]{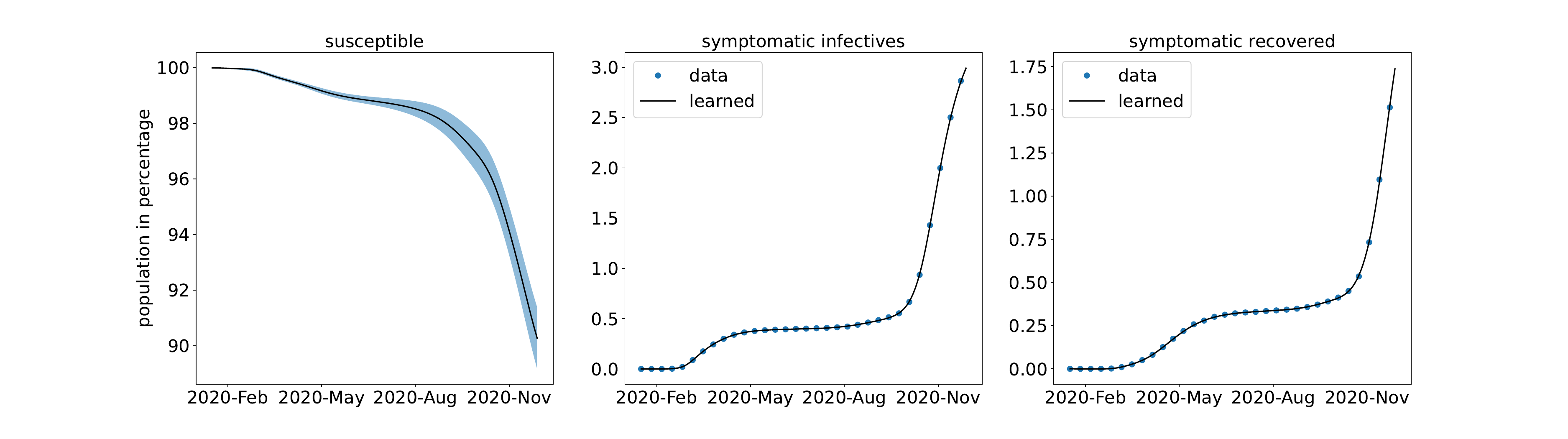}
		\caption{ }
		\label{italy_learn_params_01}	
	\end{subfigure}
	\quad %\vfill
	\begin{subfigure}[b]{\textwidth}
		\captionsetup{width=0.5\textwidth}
		\centering
		\includegraphics[width=17 cm]{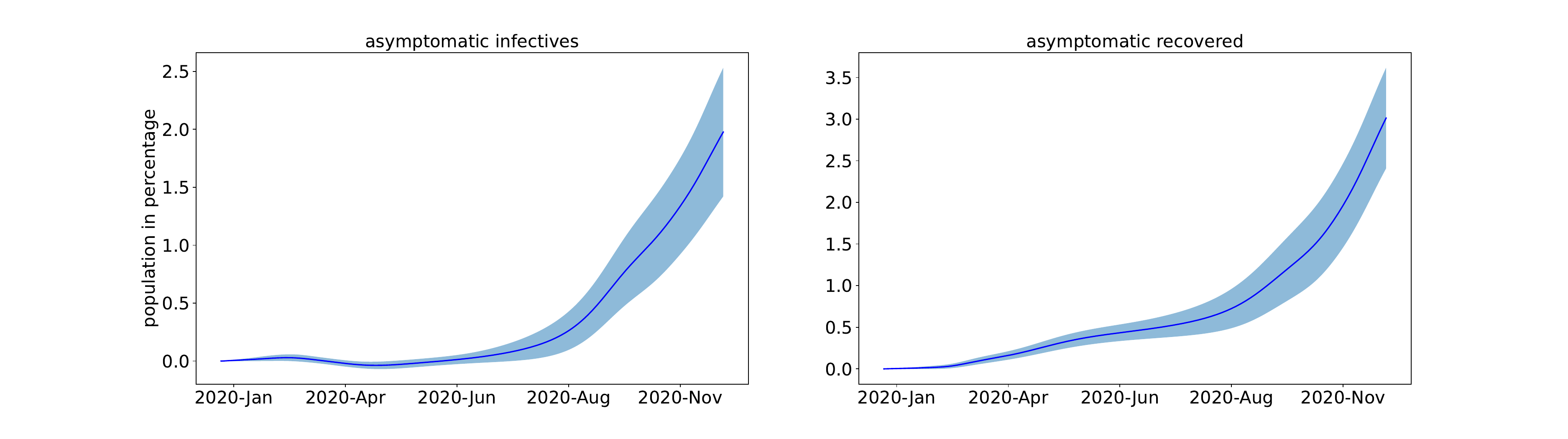}
		\caption{ }
	 	\label{italy_learn_params_02}
	\end{subfigure}
%	\quad%\vfill
%	\begin{subfigure}[b]{\textwidth}
%		\captionsetup{width=0.8\textwidth}
%		%\centering
%		\includegraphics[width=12.5 cm]{italy_relative_error}
%		\caption{relative $l_2$ error of the learned symptomatic infectives $(I)$ and recovered $(R)$}				\label{italy_learn_params_03}
%	\end{subfigure}
	\caption{Simulation of Italy COVID-19 data%Subcaption is added inside, please make sure the caption, the same as below.
	; (\textbf{a}) The learned symptomatic infectives and recovered population by the EINN Algorithm \ref{alg:Epidemiology informed neural networksird_cons}; (\textbf{b}) EINN Algorithm \ref{alg:Epidemiology informed neural networksird_cons} learns the cumulative population of Italy that are asymptomatic infectives and asymptomatic recovered from 31 January to 11 December.} \label{italy_learn_params}
\end{figure}

\begin{figure}[H]
\centering
\begin{subfigure}[b]{\textwidth}
\captionsetup{width=0.4\textwidth}
\centering
\includegraphics[width=17 cm]{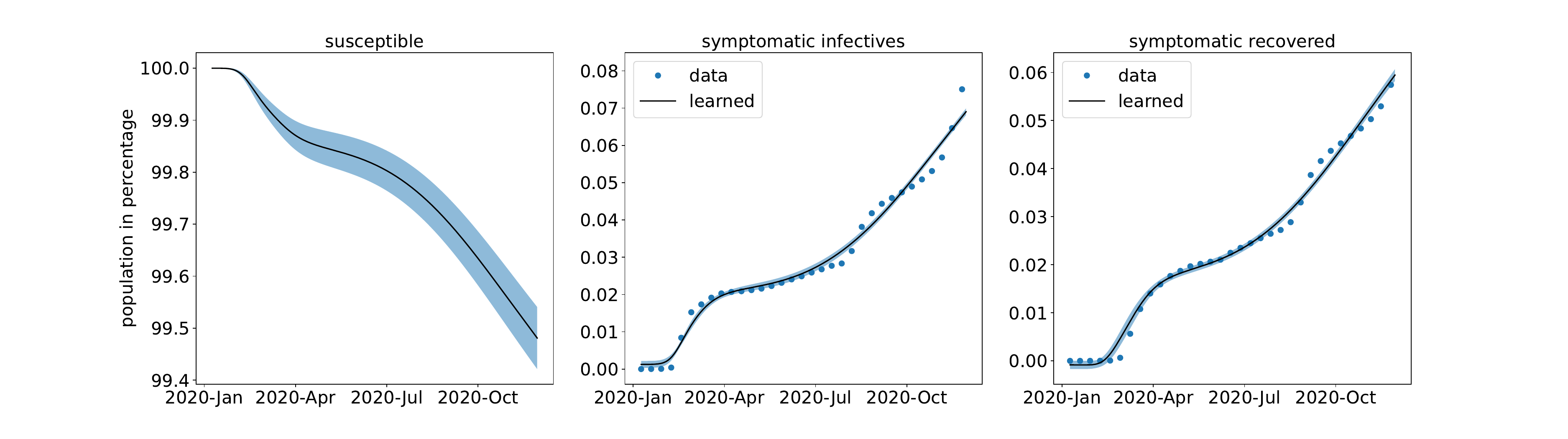}
\caption{}
 \label{skorea_learn_params_01}
\end{subfigure}
\quad %\vfill
\begin{subfigure}[b]{\textwidth}
\captionsetup{width=0.4\textwidth}
\centering
\includegraphics[width=17 cm]{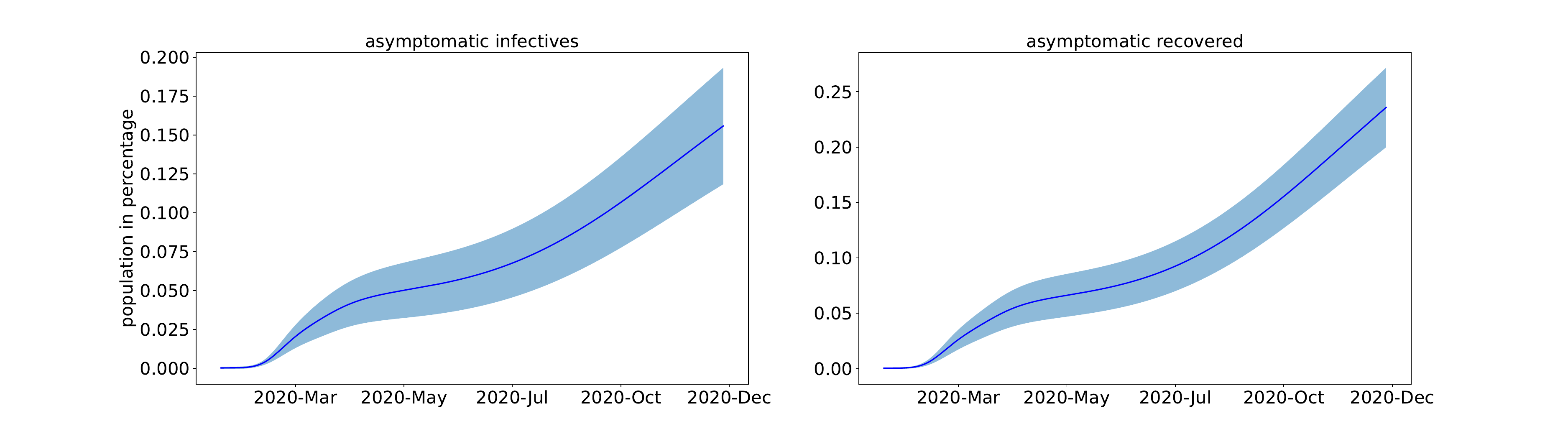}
\caption{ }
\label{skorea_learn_params_02}
\end{subfigure}
\caption{Simulation of South Korea COVID-19 data; (\textbf{a}) The learned symptomatic infectives and recovered population were obtained by the EINN Algorithm \ref{alg:Epidemiology informed neural networksird_cons}; (\textbf{b}) EINN Algorithm \ref{alg:Epidemiology informed neural networksird_cons} learns the cumulative population of South Korea that are asymptomatic infectives and asymptomatic recovered from 22 January to 11 December. 
.}
 \label{skorea_learn_params}
\end{figure}

\begin{figure}[H]
\centering
\begin{subfigure}[b]{\textwidth}
\captionsetup{width=0.4\textwidth}
\centering
\includegraphics[width=17 cm]{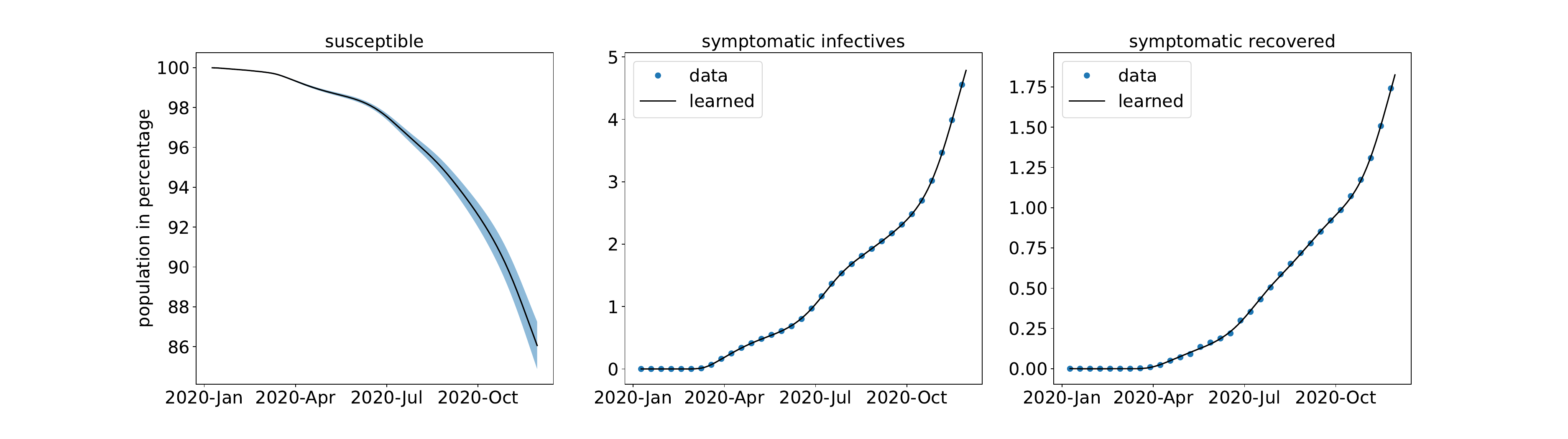}
\caption{ }
\label{usa_learn_params_01}
\end{subfigure}
\vfill
\begin{subfigure}[b]{\textwidth}
\captionsetup{width=0.4\textwidth}
\centering
\includegraphics[width=17 cm]{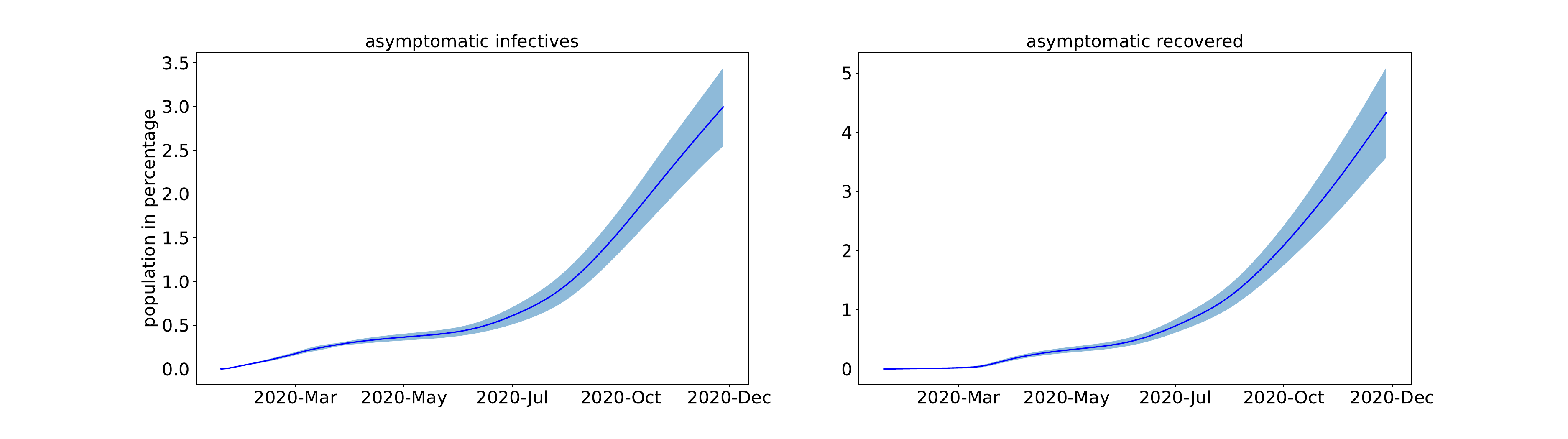}
\caption{}
\label{usa_learn_params_02}
\end{subfigure}
%\vfill
%\begin{subfigure}[b]{\textwidth}
%\captionsetup{width=0.8\textwidth}
%%\centering
%\includegraphics[width=12.5 cm]{usa_relative_error}
%\caption{relative $l_2$ error of the learned symptomatic infectives $(I)$ and recovered $(R)$. \label{usa_learn_params_03}}
%\end{subfigure}
\caption{Simulation of USA COVID-19 data; (\textbf{a}) The learned symptomatic infectives and recovered population were obtained by the EINN Algorithm \ref{alg:Epidemiology informed neural networksird_cons}; (\textbf{b}) EINN Algorithm \ref{alg:Epidemiology informed neural networksird_cons} learns the cumulative population of USA that are asymptomatic infectives and asymptomatic recovered from  22 January to 11 December.}
\label{usa_learn_params}
\end{figure}

Using data from Italy, South Korea, and the United States starting from the date of the first reported cases in the respective countries to the day before vaccination data were reported.
The cumulative infective and recovered population data are observed to be non-exponential whenever a mitigation measure such as a comprehensive lockdown is detected in the data.
We take the total population $N$ to be $60.36 \times 10^{6}$, $51.64 \times 10^{6}$, and $328.2 \times 10^{6}$ in Italy, South Korea and the USA, respectively. In Figures \ref{italy_learn_params}a--\ref{usa_learn_params}a, $M_{\kappa}$ is zero and so $\kappa = 0$ for all the period from the first reported cases to the day before vaccination data are reported.
In addition to learning the parameters, EINN learns $\xi$, the probability that an infective is reported.
High value of $\xi$ indicates large number of reported infectives.
%high reported infectives.

%%%%%%%%%%%%%%%%%%%%%%%%%%%%%%%%%%%%%%%%%%%%%%%%%%%%

%%%%%%%%%%%%%%%%%%%%%%%%%%%%%%%%%%%%%%%%%%%%%%%%%%%%

%%%%%%%%%%%%%%%%%%%%%%%%%%%%%%%%%%%%%%%%%%%%%%%%%%%%

\begin{table}[htbp]
\centering
\begin{tabular}{lcc}
%\addlinespace
\toprule
{Parameters} & {Mean} & {Std} \\
\toprule
$\beta$  & 0.03773 & 0.00276 \\
$\xi$  & 0.55699 & 0.07896  \\
$\gamma$  & 0.01327 & 0.00027 \\
$\mu$  & 0.02906 & 0.017478 \\
$\mathcal{R}_0$  & 2.32770 & 0.06014 \\
\midrule
\end{tabular}
\caption{In Figure~\eqref{italy_learn_params}, EINN Algorithm \ref{alg:Epidemiology informed neural networksird_cons} learns the constant model parameters $\beta$ $\gamma$, $\mu$, $\xi$, and $\mathcal{R}_0$ from 31 January 2020 to 11 December 2020}
\label{Table:it1}
\end{table}

\begin{table}[htbp]
\centering
\begin{tabular}{lcc}
\addlinespace
\toprule
{Parameters} & {Mean} & {Std} \\
\toprule
$\beta$  & 0.01537 & 0.00350 \\
$\xi$  & 0.24862 & 0.04333  \\
$\gamma$  & 0.00537 & 0.00013 \\
$\mu$  & 0.01174 & 0.00587 \\
$\mathcal{R}_0$  & 1.84796 & 0.16187 \\
\midrule
\end{tabular}
\caption{In Figure~\eqref{skorea_learn_params}, EINN Algorithm \ref{alg:Epidemiology informed neural networksird_cons} learns the constant model parameters $\beta$ $\gamma$, $\mu$, $\xi$, and $\mathcal{R}_0$ from 22 January 2020 to 11 December 2020}
\label{Table:sk1}
\end{table}

\begin{table}[htbp]
\centering
\begin{tabular}{lcc}
\addlinespace
\toprule
{Parameters} & {Mean} & {Std} \\
\toprule
$\beta$  & 0.02130 & 0.00144 \\
$\xi$  & 0.49176 & 0.06541  \\
$\gamma$  & 0.00437 & 0.000046 \\
$\mu$  & 0.01499 & 0.00199 \\
$\mathcal{R}_0$  & 3.10406 & 0.09609 \\
\midrule
\end{tabular}
\caption{In Figure~\eqref{usa_learn_params}, EINN Algorithm \ref{alg:Epidemiology informed neural networksird_cons} learns the constant model parameters $\beta$ $\gamma$, $\mu$, $\xi$, and $\mathcal{R}_0$ from 22 January 2020 to 11 December 2020}
\label{Table:us1}
\end{table}

As shown in Figures~\eqref{italy_learn_params}a--\eqref{usa_learn_params}a, early in the pandemic, the cumulative infective and recovered data closely resemble an exponential function. Cubic Spline interpolation is used to generate 3000 training points from the cumulative symptomatic infective and recovered data. In Tables~\eqref{Table:it1}--\eqref{Table:us1} the mean and standard deviation of the the parameters $\beta$ $\gamma$, $\mu$, $\xi$, and $\mathcal{R}_0$ are presented after 10 runs of EINN Algorithm~\eqref{alg:Epidemiology informed neural networksird_cons}

\subsection{Data-Driven Simulation Results for Non-Pharmaceutical Mitigation Measures} 
\label{cons_Mit}

The model parameters in an epidemiology model are influenced by mitigation measures.
For instance, social distancing corresponds to reducing the transmission rate by reducing human contact.
In this Section, we simulate different levels of various non-pharmaceutical mitigation measure, and we demonstrate their impact on $\mathcal{R}_0$ and the spread of COVID-19.

\subsubsection{Early Detection of Infectives} \label{cparam}
%%%%%%%%%%%%%%%%%%%%%%%%%%%%%%%%%%%%%%%%%%%%%%%%%%%%%%%%%%%%%%%%%%%%%%%%%%%%%

Early detection of infectives population leads to higher reported infectives. 
This results in an early isolation of individuals who have had contact with infective individuals.
There are no reported data for the asymptomatic infectives populations. 
Simulating with higher $\xi$ increases the symptomatic infectives population. 
This corresponds to higher reported cases. 
Simulations are presented for Italy, South Korea, and the USA see Tables~\ref{Table:1}--\ref{Table:2b}.

\begin{table}[htbp]
%\footnotesize
\centering
\begin{tabular}{rrcccccc}
\addlinespace
\toprule
\multicolumn{ 2}{c}{{\bf }} & {\bf $\beta$} & {\bf $\gamma$} & {\bf $\mu$} & {\bf $\beta \xi$} & {\bf $\beta(1-\xi)$} & {\bf $\mathcal{R}_0$}\\
\toprule
{\bf $\xi = 0.1$ } & {\bf Mean} & 0.03161 & 0.00119 & 0.03125 & 0.00316 & 0.02845 & 4.32459 \\
{\bf } & {\bf Std} & 0.00376 & 0.00047 & 0.02510 & 0.00038 & 0.00338 & 0.96941 \\
\midrule
{\bf $\xi = 0.25$ } & {\bf Mean} & 0.03827 & 0.00810 & 0.02418 & 0.00957 & 0.02870 & 2.42050 \\
{\bf } & {\bf Std} & 0.00307 & 0.00122 & 0.00456 & 0.00077 & 0.00230 & 0.08582 \\
\midrule
{\bf $\xi = 0.50$ } & {\bf Mean} & 0.03698 & 0.01208 & 0.03253 & 0.01849 & 0.01849 & 2.33102 \\
{\bf } & {\bf Std} & 0.00304 & 0.00152 & 0.02876 & 0.00152 & 0.00152 & 0.11068 \\
\midrule
{\bf $\xi = 0.75$ } & {\bf Mean} & 0.03700 & 0.01435 & 0.03027 & 0.02775 & 0.00925 & 2.35074 \\
{\bf } & {\bf Std} & 0.00262 & 0.00122 & 0.01801 & 0.00196 & 0.00065 & 0.09155 \\
\midrule
\end{tabular}
\caption{The learned parameters using EINN Algorithm \ref{alg:Epidemiology informed neural networksird_cons} with fixed values of $\xi$ based on Italy data from 31 January 2020 to 5 September 2020.}
\label{Table:1}
\end{table}

\begin{table}[htbp]
\centering
\begin{tabular}{rrcccccc}
\addlinespace
\toprule
\multicolumn{ 2}{c}{{\bf }} & {\bf $\beta$} & {\bf $\gamma$} & {\bf $\mu$} & {\bf $\beta \xi$} & {\bf $\beta(1-\xi)$} & {\bf $\mathcal{R}_0$}\\
\toprule
{\bf $\xi = 0.1$ } & {\bf Mean} & 0.01230 & 0.00179 & 0.00958 & 0.00123 & 0.01107 & 1.95802 \\
{\bf } & {\bf Std} & 0.00172 & 0.00041 & 0.00304 & 0.00017 & 0.00155 & 0.15387 \\
\midrule
{\bf $\xi = 0.25$ } & {\bf Mean} & 0.01326 & 0.00615 & 0.00792 & 0.00332 & 0.00995 & 1.83806 \\
{\bf } & {\bf Std} & 0.00101 & 0.00118 & 0.00148 & 0.00025 & 0.00076 & 0.11778 \\
\midrule
{\bf $\xi = 0.50$ } & {\bf Mean} & 0.01499 & 0.01222 & 0.00754 & 0.00749 & 0.00749 & 1.73132 \\
{\bf } & {\bf Std} & 0.00217 & 0.00156 & 0.00269 & 0.00109 & 0.00109 & 0.22998 \\
\midrule
{\bf $\xi = 0.75$ } & {\bf Mean} & 0.01195 & 0.01537 & 0.00318 & 0.00896 & 0.00299 & 1.64407 \\
{\bf } & {\bf Std} & 0.00186 & 0.00226 & 0.00224 & 0.00139 & 0.00047 & 0.28103 \\
\midrule
\end{tabular}
\caption{The learned parameters using EINN Algorithm \ref{alg:Epidemiology informed neural networksird_cons} with fixed values of $\xi$ based on South Korea data from 22 January 2020 to 5 September 2020.}
\label{Table:2}
\end{table}

\begin{table}[htbp]
\centering
\begin{tabular}{rrcccccc}
\addlinespace
\toprule
\multicolumn{ 2}{c}{{\bf }} & {\bf $\beta$} & {\bf $\gamma$} & {\bf $\mu$} & {\bf $\beta \xi$} & {\bf $\beta(1-\xi)$} & {\bf $\mathcal{R}_0$}\\
\toprule
{\bf $\xi = 0.25$ } & {\bf Mean} & 0.02270 & 0.00224 & 0.01471 & 0.00568 & 0.01703 & 3.83612 \\
{\bf } & {\bf Std} & 0.00143 & 0.00056 & 0.00119 & 0.00036 & 0.00108 & 0.53227 \\
\midrule
{\bf $\xi = 0.50$ } & {\bf Mean} & 0.02126 & 0.00419 & 0.01639 & 0.01063 & 0.01063 & 3.20680 \\
{\bf } & {\bf Std} & 0.00071 & 0.00032 & 0.00239 & 0.00036 & 0.00036 & 0.13379\\
\midrule
{\bf $\xi = 0.75$ } & {\bf Mean} & 0.02009 & 0.00514 & 0.02039 & 0.01507 & 0.00502 & 3.18964 \\
{\bf } & {\bf Std} & 0.00083 & 0.00026 & 0.00465 & 0.00062 & 0.00021 & 0.09912 \\
\midrule
\end{tabular}
\caption{The learned parameters using EINN Algorithm \ref{alg:Epidemiology informed neural networksird_cons} with fixed values of $\xi$ based on USA data from 22 January 2020 to 5 September 2020.}
\label{Table:2b}
\end{table}

Higher $\xi$ values in Tables~\ref{Table:1}--\ref{Table:2b}, increase the symptomatic infectives population and reduce the asymptomatic population in general. 
This is reflected by the increase in the  $\beta \xi$ column and the corresponding decrease in the $\beta(1-\xi)$ column.
This means that more people will be in hospitalization/isolation. 
This translates to more recovery in the symptomatic compartment.
We see that the detection of early infectives alone is not enough to mitigate an infectious disease such as COVID-19 as demonstrated in the $\mathcal{R}_0$ column in Tables~\ref{Table:1}--\ref{Table:2b}. 
It should be combined with other measures such as contact tracing of infectives.

%Higher $\xi$ values in Tables~\ref{Table:1}--\ref{Table:2b}, increase the symptomatic infectives population and reduce the asymptomatic population in general. 
%This is reflected by the increase in the  $\beta \xi$ column and the corresponding decrease in the $\beta(1-\xi)$ column.
%This means that more people will be in hospitalization/isolation. 
%This translates to more recovery in the symptomatic compartment.
%We see that the detection of early infectives alone is not enough to mitigate an infectious disease such as COVID-19. 
%It should be combined with other measures such as contact tracing of infectives.
%It is assumed that the total infective population is unchanged as we raise $\xi$. 
%$\mathcal{R}_0$ decreases in general.
%If the asymptomatic infectives population is much higher than the symptomatic population, this results in higher $\mathcal{R}_0$. 
%There is a right balance between $I$ and $J$ population.  

%%%%%%%%%%%%%%%%%%%%%%%%%%%%%%%%%%%%%%%%%%%%%%%%%%%%%%%%%%%%%%%%%%%%%%%%%%%%%
\subsubsection{Social Distancing} \label{soc_dis}
%%%%%%%%%%%%%%%%%%%%%%%%%%%%%%%%%%%%%%%%%%%%%%%%%%%%%%%%%%%%%%%%%%%%%%%%%%%%%

It is widely understood that measures such as a lockdown, social distancing, and widespread adoption of facial coverings result in the mitigation of COVID-19. 
Social distancing is often the most sought-after measure at reducing the $\mathcal{R}_0$. 
The goal of social distancing is to reduce the average number of human contacts. 
This is demonstrated by reducing $\beta$, the transmission rate \cite{Gaeta2021}.
The impact of social distancing on the $\mathcal{R}_0$ is presented in the following Tables~\ref{Table:7}--\ref{Table:8b}.

\begin{table}[htbp]
\centering
\begin{tabular}{rrcccccc}
\addlinespace
\toprule
\multicolumn{ 2}{c}{{\bf }} & {\bf $\gamma$} & {\bf $\xi$} & {\bf $\mu$} & {\bf $\beta \xi$} & {\bf $\beta(1-\xi)$} & {\bf $\mathcal{R}_0$}\\
\toprule
{\bf $\beta = 0.1$ } & {\bf Mean} & 0.01371 & 0.69461 & 0.36808 & 0.06946 & 0.03054 & 5.15392 \\
{\bf } & {\bf Std} & 0.00020 & 0.04506 & 0.07684 & 0.00451 & 0.00451 & 0.24303 \\
\midrule
{\bf $\beta = 0.05$ } & {\bf Mean} & 0.01361 & 0.55675 & 0.30860 & 0.02784 & 0.02216 & 2.11654 \\
{\bf } & {\bf Std} & 0.00000 & 0.00000 & 0.14114 & 0.00000 & 0.00000 & 0.00000 \\
\midrule
{\bf $\beta = 0.025$ } & {\bf Mean} & 0.01163 & 0.54429 & 0.021192 & 0.01361 & 0.01139 & 1.81348 \\
{\bf } & {\bf Std} & 0.00032 & 0.09003 & 0.01164 & 0.00225 & 0.00225 & 0.45399 \\
\midrule
% {$\bf \beta = 0.01$ } & {\bf Mean} & 0.00726 & 0.53089 & 0.00341 & 0.00531 & 0.00469 & 3.76097 \\
% {\bf } & {\bf Std} & 0.00016 & 0.02624 & 0.00238 & 0.00026 & 0.00026 & 2.12097 \\
% \midrule
\end{tabular}
\caption{The learned parameters using EINN Algorithm \ref{alg:Epidemiology informed neural networksird_cons} with fixed values of $\beta$ based on Italy data from 31 January 2020 to 5 September 2020}
\label{Table:7}
\end{table}

\begin{table}[htbp]
\centering
\begin{tabular}{rrcccccc}
\addlinespace
\toprule
\multicolumn{ 2}{c}{{\bf }} & {\bf $\gamma$} & {\bf $\xi$} & {\bf $\mu$} & {\bf $\beta \xi$} & {\bf $\beta(1-\xi)$} & {\bf $\mathcal{R}_0$}\\
\toprule
% {$\bf \beta = 0.1$ } & {\bf Mean} & 0.0111 & 0.0013 & 0.0343 & 0.0011 & 0.0099 & 0.3578 \\
% {\bf } & {\bf Std} & 0.0111 & 0.0013 & 0.0343 & 0.0011 & 0.0099 & 0.3578 \\
% \midrule
{\bf $\beta = 0.05$ } & {\bf Mean} & 0.00571 & 0.64985 & 0.16848 & 0.03249 & 0.01751 & 7.69632 \\
{\bf } & {\bf Std} & 0.00229 & 0.18684 & 0.03398 & 0.00934 & 0.00934 & 4.61209 \\
\midrule
{\bf $\beta = 0.025$ } & {\bf Mean} & 0.00539 & 0.30717 & 0.03705 & 0.00768 & 0.01732 & 1.90597 \\
{\bf } & {\bf Std} & 0.00000 & 0.00000 & 0.01178 & 0.00000 & 0.00000 & 0.00000 \\
\midrule
{\bf $\beta = 0.01$ } & {\bf Mean} & 0.00285 & 0.09468 & 0.00819 & 0.00095 & 0.00905 & 1.37149 \\
{\bf } & {\bf Std} & 0.00103 & 0.07145 & 0.002224 & 0.00071 & 0.00071 & 0.48262 \\
\midrule
\end{tabular}
\caption{The learned parameters using EINN Algorithm \ref{alg:Epidemiology informed neural networksird_cons} with fixed values of $\beta$ based on South Korea data from 22 January 2020 to 5 September 2020}
\label{Table:8}
\end{table}

\begin{table}[htbp]
\centering
\begin{tabular}{rrcccccc}
\addlinespace
\toprule
\multicolumn{ 2}{c}{{\bf }} & {\bf $\gamma$} & {\bf $\xi$} & {\bf $\mu$} & {\bf $\beta \xi$} & {\bf $\beta(1-\xi)$} & {\bf $\mathcal{R}_0$}\\
\toprule
% {$\bf \beta = 0.1$ } & {\bf Mean} & 0.00458 & 0.51246 & 0.0343 & 0.0011 & 0.0099 & 0.3578 \\
% {\bf } & {\bf Std} & 0.00005 & 0.06541 & 0.0343 & 0.0011 & 0.0099 & 0.3578 \\
% \midrule
{\bf $\beta = 0.05$ } & {\bf Mean} & 0.00398 & 0.32026 & 0.64670 & 0.01601 & 0.03399 & 4.53989 \\
{\bf } & {\bf Std} & 0.00115 & 0.05013 & 0.17882 & 0.00251 & 0.00251 & 1.61048 \\
\midrule
{\bf $\beta = 0.025$ } & {\bf Mean} & 0.00458 & 0.85867 & 0.03475 & 0.02147 & 0.00353 & 3.34950 \\
{\bf } & {\bf Std} & 0.00048 & 1.22557 & 0.01843 & 0.03064 & 0.03064 & 1.52165 \\
\midrule
{\bf $\beta = 0.01$ } & {\bf Mean} & 0.00314 & 0.59924 & 0.00532 & 0.00599 & 0.00401 & 3.52616 \\
{\bf } & {\bf Std} & 0.00009 & 0.21138 & 0.00613 & 0.00211 & 0.00211 & 1.52799 \\
\midrule
\end{tabular}
\caption{The learned parameters using EINN Algorithm \ref{alg:Epidemiology informed neural networksird_cons} with fixed values of $\beta$ based on USA data from 22 January 2020 to 5 September 2020}
\label{Table:8b}
\end{table}

Reducing $\beta$ in Tables~\ref{Table:7}--\ref{Table:8b} correspond to a reduced symptomatic infectives population $I$. 
%However, the total infectives are unchanged as we reduce $\beta$. 
There is an increase in the asymptomatic infectives population $J$. 
%$\mathcal{R}_0$ does not decrease in general. 
Social distancing is effective when the asymptomatic infective population $J$ diminishes. 
$\beta\xi$ and $\beta(1-\xi)$ both decreases. 
Social distancing should be combined with contact tracing and early detection of infectives~population.

%%%%%%%%%%%%%%%%%%%%%%%%%%%%%%%%%%%%%%%%%%%%%%%%%%%%%%%%%%%%%%%%%%%%%%%%%%%%%
\subsubsection{Contact Tracing of Infectives} \label{ctrace}
%%%%%%%%%%%%%%%%%%%%%%%%%%%%%%%%%%%%%%%%%%%%%%%%%%%%%%%%%%%%%%%%%%%%%%%%%%%%%

Contact tracing is equivalent to increasing the symptomatic recovery and asymptomatic recovery rates \cite{Gaeta2021}. However, since we do not have reported data for the asymptomatic population, in this paper, we pursue contact tracing as an increase in the symptomatic recovery rate. 
This is equivalent to reducing the number of days an infective individual stays infective. 
In Tables~\ref{Table:4}--\ref{Table:6}, the impact of contact tracing is demonstrated by increasing the symptomatic recovery rate.

\begin{table}[htbp]
\centering
\begin{tabular}{rrcccccc}
\addlinespace
\toprule
\multicolumn{ 2}{c}{{\bf }} & {\bf $\beta$} & {\bf $\xi$} & {\bf $\mu$} & {\bf $\beta \xi$} & {\bf $\beta(1-\xi)$} & {\bf $\mathcal{R}_0$}\\
\toprule
{\bf $\gamma = 0.001$ } & {\bf Mean} & 0.03235 & 0.37063 & 0.02157 & 0.01208 & 0.02027 & 13.05386 \\
{\bf } & {\bf Std} & 0.00251 & 0.05151 & 0.00443 & 0.00245 & 0.00127 & 2.29273 \\
\midrule
{\bf $\gamma = 0.005$ } & {\bf Mean} & 0.03386 & 0.43284 & 0.02479 & 0.01484 & 0.01902 & 3.86667 \\
{\bf } & {\bf Std} & 0.00372 & 0.08534 & 0.01229 & 0.00406 & 0.00258 & 0.50726 \\
\midrule
{\bf $\gamma = 0.01$ } & {\bf Mean} & 0.03564 & 0.49312 & 0.02306 & 0.01771 & 0.01793 & 2.62805 \\
{\bf } & {\bf Std} & 0.00332 & 0.07613 & 0.00791 & 0.00373 & 0.00216 & 0.12344 \\
\midrule
{\bf $\gamma = 0.05$ } & {\bf Mean} & 0.04573 & 0.85962 & 0.01113 & 0.03924 & 0.00649 & 1.36939 \\
{\bf } & {\bf Std} & 0.00223 & 0.06479 & 0.00436 & 0.00270 & 0.00319 & 0.09824 \\
\midrule
\end{tabular}
\caption{The learned parameters using EINN Algorithm \ref{alg:Epidemiology informed neural networksird_cons} with fixed values of $\gamma$ based on Italy data from 31 January 2020 to 5 September 2020}
\label{Table:4}
\end{table}

\begin{table}[htbp]
\centering
\begin{tabular}{rrcccccc}
\addlinespace
\toprule
\multicolumn{ 2}{c}{{\bf }} & {\bf $\beta$} & {\bf $\xi$} & {\bf $\mu$} & {\bf $\beta \xi$} & {\bf $\beta(1-\xi)$} & {\bf $\mathcal{R}_0$}\\
\toprule
{\bf $\gamma = 0.001$ } & {\bf Mean} & 0.01274 & 0.17877 & 0.00811 & 0.00225 & 0.01049 & 3.59469 \\
{\bf } & {\bf Std} & 0.00161 & 0.02628 & 0.00199 & 0.00027 & 0.00153 & 0.19821 \\
\midrule
{\bf $\gamma = 0.005$ } & {\bf Mean} & 0.01386 & 0.24278 & 0.00890 & 0.00335 & 0.01051 & 1.90717 \\
{\bf } & {\bf Std} & 0.00157 & 0.02331 & 0.00261 & 0.00039 & 0.00134 & 0.17414 \\
\midrule
{\bf $\gamma = 0.01$ } & {\bf Mean} & 0.01410 & 0.27970 & 0.00841 & 0.00399 & 0.01012 & 1.74857 \\
{\bf } & {\bf Std} & 0.00239 & 0.02634 & 0.00350 & 0.00097 & 0.00149 & 0.30925 \\
\midrule
{\bf $\gamma = 0.05$ } & {\bf Mean} & 0.01804 & 0.69863 & 0.00552 & 0.01269 & 0.00534 & 1.31972 \\
{\bf } & {\bf Std} & 0.00309 & 0.08219 & 0.00234 & 0.00311 & 0.00137 & 0.27368 \\
\midrule
\end{tabular}
\caption{The learned parameters using EINN Algorithm \ref{alg:Epidemiology informed neural networksird_cons} with fixed values of $\gamma$ based on South Korea data from 22 January 2020 to 5 September 2020}
\label{Table:5}
\end{table}

\begin{table}[htbp]
\centering
\begin{tabular}{rrcccccc}
\addlinespace
\toprule
\multicolumn{ 2}{c}{{\bf }} & {\bf $\beta$} & {\bf $\xi$} & {\bf $\mu$} & {\bf $\beta \xi$} & {\bf $\beta(1-\xi)$} & {\bf $\mathcal{R}_0$}\\
\toprule
{\bf $\gamma = 0.001$ } & {\bf Mean} & 0.02089 & 0.38780 & 0.01533 & 0.00808 & 0.01281 & 8.92979 \\
{\bf } & {\bf Std} & 0.00091 & 0.04234 & 0.00219 & 0.00073 & 0.00127 & 0.60324 \\
\midrule
{\bf $\gamma = 0.005$ } & {\bf Mean} & 0.02121 & 0.50579 & 0.01446 & 0.01071 & 0.01049 & 2.88515 \\
{\bf } & {\bf Std} & 0.00105 & 0.04284 & 0.00240 & 0.00084 & 0.00120 & 0.07506 \\
\midrule
{\bf $\gamma = 0.01$ } & {\bf Mean} & 0.02334 & 0.56126 & 0.01437 & 0.01308 & 0.01026 & 2.02548 \\
{\bf } & {\bf Std} & 0.00087 & 0.03675 & 0.00156 & 0.00069 & 0.00117 & 0.02210 \\
% \midrule
% {$\bf \gamma = 0.05$ } & {\bf Mean} & 0.0111 & 0.0013 & 0.0343 & 0.0011 & 0.0099 & 0.3578 \\
% {\bf } & {\bf Std} & 0.0111 & 0.0013 & 0.0343 & 0.0011 & 0.0099 & 0.3578 \\
\midrule
\end{tabular}
\caption{The learned parameters using EINN Algorithm \ref{alg:Epidemiology informed neural networksird_cons} with fixed values of $\gamma$ based on USA data from 22 January 2020 to 5 September 2020}
\label{Table:6}
\end{table}

The raising of $\gamma$ in Tables~\ref{Table:4}--\ref{Table:6}, increases the symptomatic infectives population $I$ which is demonstrated in increased $\xi$ and increased $\beta$. $\beta(1-\xi)$ decreases while $\beta \xi$ increases. 
This also results in a reduced $\mathcal{R}_0$. 
Contact tracing is an efficient mitigation measure in lowering the spread of COVID-19.

%The raising of $\gamma$ in Tables~\ref{Table:4}--\ref{Table:6}, increases the symptomatic infectives population $I$ which is demonstrated in increased $\xi$ and increased $\beta$. $\beta(1-\xi)$ decreases while $\beta \xi$ increases. 
%This also results in a reduced $\mathcal{R}_0$. 
%Contact tracing is an efficient mitigation measure in lowering the spread of COVID-19.

%%%%%%%%%%%%%%%%%%%%%%%%%%%%%%%%%%%%%%%%%%%%%%%%%%%%%%%%%%%%%%%%%%%%%%%%%%%%%
\subsection{Data-Driven Simulation Results for Vaccination Efficacy} 
\label{vaccination}

\begin{figure}[htbp]
\centering
\begin{subfigure}[H]{\textwidth}
\captionsetup{width=0.4\textwidth}
\centering
\includegraphics[width=14 cm]{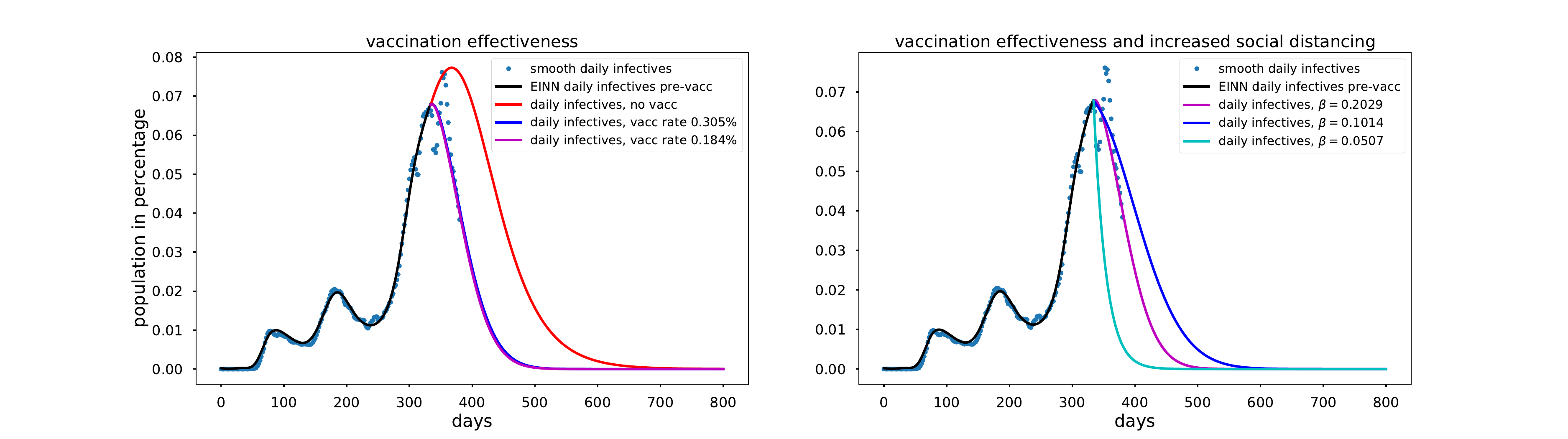} 
\end{subfigure}
\vspace{+6pt}
\quad %\vfill
\begin{subfigure}[H]{\textwidth}
\captionsetup{width=0.4\textwidth}
\centering
\includegraphics[width=14 cm]{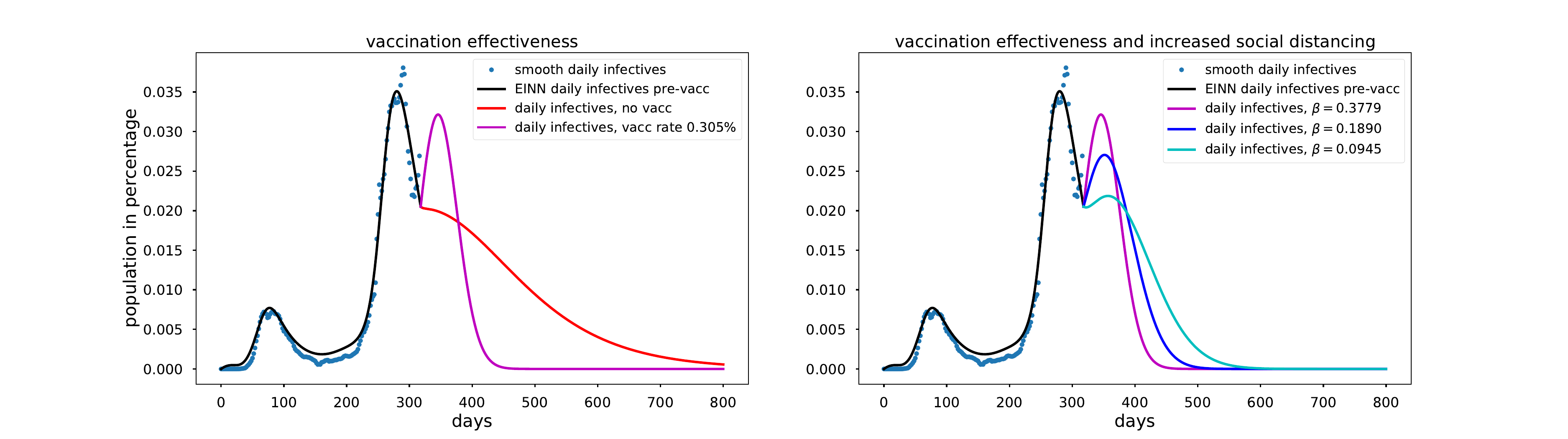} 
\end{subfigure}
\caption{Vaccination efficacy}
\label{us_uk_vac}
\end{figure}

The mitigation measures described in Section \ref{cons_Mit} are non-pharmaceutical measures. 
In this Section, we discuss vaccination.
In the fight against COVID-19, countries such as USA and United Kingdom began to vaccinate in December 2020. 
A major goal of vaccination is to reduce the susceptible population, i.e., people recover without becoming infected. 
This constitutes a pharmaceutical mitigation measure. 
We considered the vaccination data for the USA and the United Kingdom, and simulate the effectiveness of vaccination on the daily reported infectives. A hybrid neural network is used to simulate an efficient vaccination strategy in~\citep*{Torku2021}. We show that the implementation of
Algorithm~\ref{alg:Epidemiology informed neural networksird_cons} for the asymptomatic-SIR model~\eqref{Asymptomatic-SIRD_0}, we can demonstrate the efficacy of vaccination in combination with some mitigation measures. In Figure~\ref{us_uk_vac} we present a simulation of the effectiveness of vaccination in combination with an increase in social distancing in the USA and in the United Kingdom.

%\begin{figure}[htbp]
%    \subfloat[UK]{\includegraphics[width = 5in]{sm-daiyData_vac_UK_01.pdf}} 
%    \vspace{+4pt}
%    \vfill
%    \subfloat[USA]{\includegraphics[width = 5in]{sm-daiyData_vac_USA_01.pdf}}
%\caption{Vaccination efficacy}
%\label{us_uk_vac}
%\end{figure}
We used the USA projection of $1,000,000$ daily vaccination. In the case of the magenta curve, we learned $\kappa$ using the daily vaccination data. The first reported case was $01/22/2020$, Vaccination data were first reported on 19 December 2020. In~\ref{us_uk_vac}(\textbf{a}) the model is extrapolated for 2 cases. The red curve is the case of no vaccination, here $\kappa = 0$. In the magenta curve, we learned $\kappa$ using the daily vaccination data. The first reported case was on 31 January 2020, Vaccination data were first reported on 13 December 2020. The effectiveness of vaccination is demonstrated by learning the pre-vaccination and post-vaccination epidemiology parameters using smooth daily reported infectious data from the USA. In~\ref{us_uk_vac}(\textbf{b}) the effectiveness of vaccination is demonstrated by learning the pre-vaccination and post-vaccination epidemiology parameters using smooth daily reported infectives data from the United Kingdom.

\subsection{Data-Driven Simulation Results for Time-Varying Transmission Rate} 
\label{gwebb_result}
\begin{figure}[htbp]
\centering
    \subfloat[Italy]{\includegraphics[width = 6in]{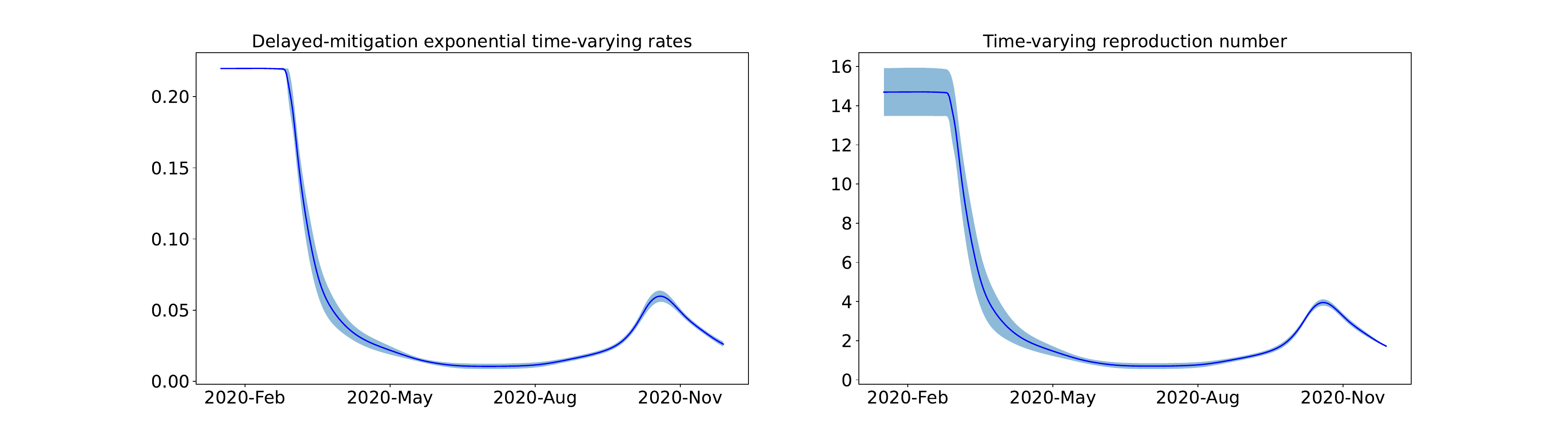}} 
    \vfill
    \subfloat[USA]{\includegraphics[width = 6in]{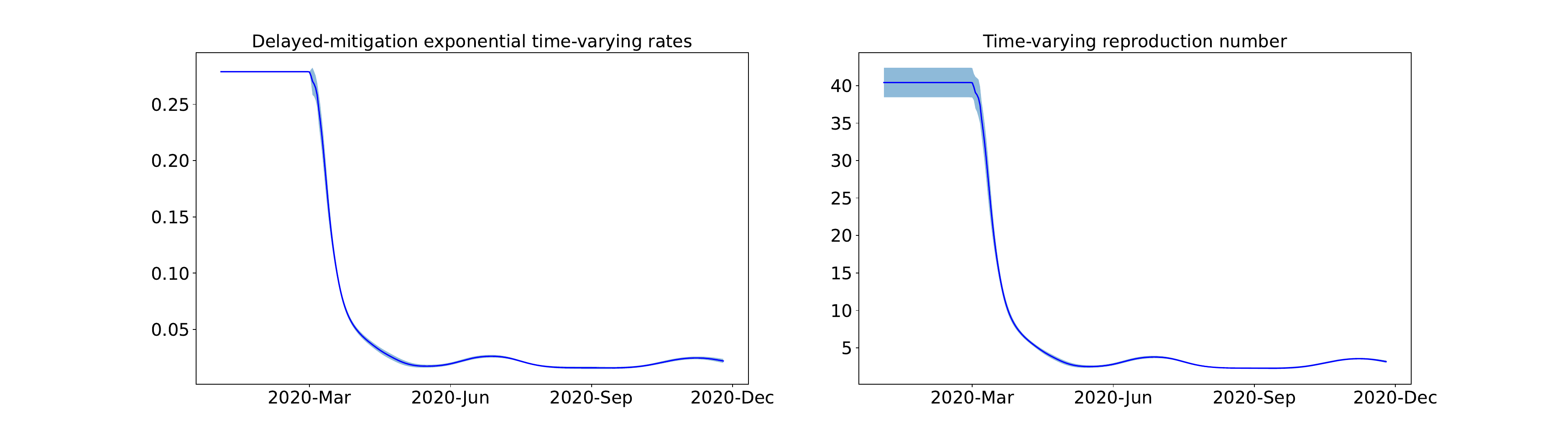}}
\caption{Delayed-mitigation exponential time-varying rates.}
\label{gWebb_fig}
\end{figure}

%\noindent 
In the EINN Algorithm~\ref{alg:Epidemiology informed neural network_gwebb}, $M_{\beta}$ corresponds to the number of days mitigation is delayed in the data, which is equal to $K$ in Equation \eqref{webb_eqn}. $M_{\kappa}$ is the number of vaccination~days. In Figures~\ref{gWebb_fig}(\textbf{a}) and~\ref{gWebb_fig}(\textbf{b}), time-varying transmission rates learned by the EINN Algorithm~\ref{alg:Epidemiology informed neural network_gwebb}.

In Figure~\ref{gWebb_fig}(\textbf{a}) A learned delayed-mitigation exponential time-varying transmission rate $\beta$ is plotted for cumulative Italy COVID-19 data from January 31, 2020 to December 11, 2020. $N = 60.36 \times 10^6$. The plotted time-varying basic reproduction rate $\mathcal{R}_t$ shows the impact of lockdown and the mitigation measures post-lockdown. The relaxation that followed is due to the COVID-19 surge and is detected in the learned $\beta$ and $\mathcal{R}_t$. The EINN Algorithm \ref{alg:Epidemiology informed neural network_gwebb} also learns $\gamma = 0.0121$ and $\mu = 0.0106$. The MSE in $(I)$ is $7.5 \times 10^{-5}$. In Figure~\ref{gWebb_fig}(\textbf{b})  A learned delayed-mitigation  exponential time-varying transmission rate $\beta$ is plotted for cumulative U.S.A COVID-19 data from 22 January 2020 to December 11, 2020. $N = 328.2 \times 10^6$. The time-varying basic reproduction rate $R_t$ is underestimated pre-lockdown. The EINN Algorithm~\ref{alg:Epidemiology informed neural network_gwebb} also learns $\gamma = 0.001$ and $\mu = 0.0224$. The MSE in $(I)$ is $3.88 \times 10^{-4}$. In~\ref{gWebb_fig}(\textbf{a}) The delayed-mitigation exponential transmission rate is learned using  Equation \eqref{webb_eqn} in Equation~\eqref{Asymptomatic-SIRD_0}. We set $K=40$ and we and fix $\xi = 0.37$ in EINN Algorithm \ref{alg:Epidemiology informed neural network_gwebb}. We take $\beta_0=0.22$, obtained using early data and nonlinear regression. EINN Algorithm \ref{alg:Epidemiology informed neural network_gwebb} learns $\eta = 0.87$, the rate at which human contact decreases. In~\ref{gWebb_fig}(\textbf{b}) The delayed-mitigation exponential transmission rate is learned using  Equation \eqref{webb_eqn} in Equation~\eqref{Asymptomatic-SIRD_0}. We set $K=57$ and we fix $\xi = 0.46$ in EINN Algorithm~\ref{alg:Epidemiology informed neural network_gwebb}. We take $\beta_0=0.279$, obtained using early data and nonlinear regression. EINN Algorithm~\ref{alg:Epidemiology informed neural network_gwebb} learns $\eta = 0.60$, the rate at which human contact decreases.

\subsection{Data-Driven Simulation for Piecewise Transmission Rate}
\label{gaeta_result}
\begin{figure}[htbp]
\centering
    \subfloat[Italy]{\includegraphics[width = 6in]{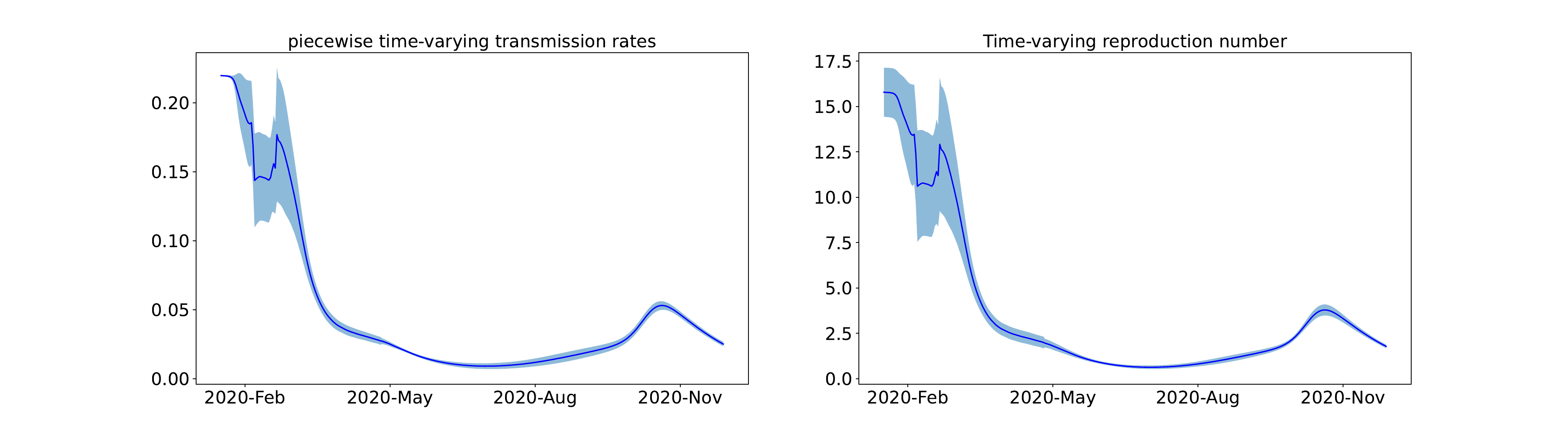}} 
    \vfill
    \subfloat[USA]{\includegraphics[width = 6in]{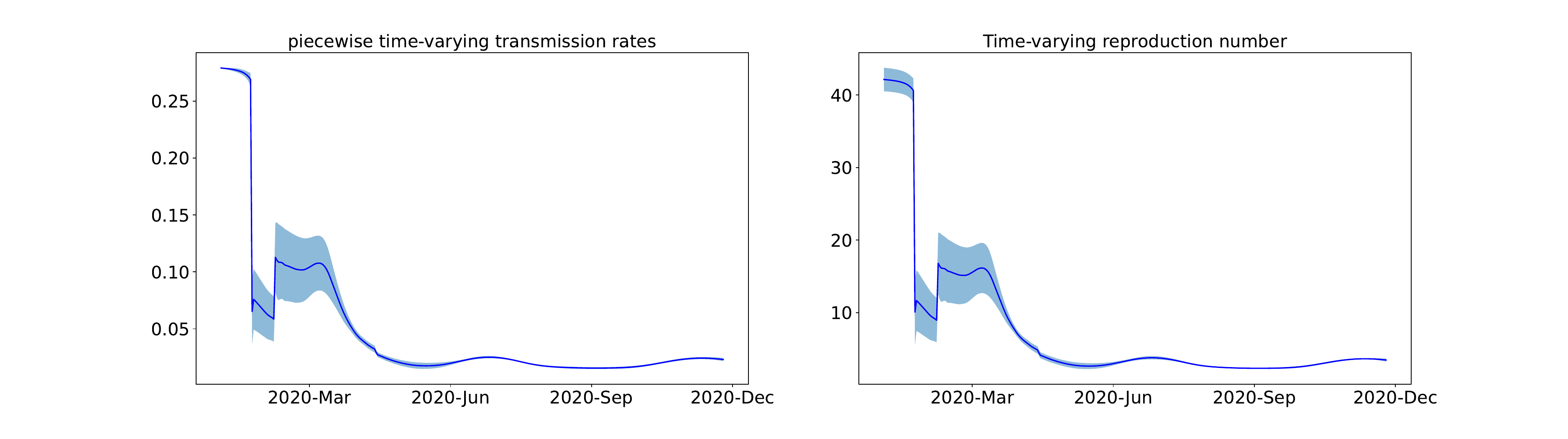}}
\caption{Piecewise-constant time-varying rates.}
\label{gaeta_fig}
\end{figure}

%\noindent 
In the EINN Algorithm~\ref{alg:Epidemiology informed neural network_gaeta}, $M_{i}, 1 \leq i \leq n$ are chosen to corresponds to a partitioning in the data. Time-varying transmission rates learned by the EINN Algorithm~\ref{alg:Epidemiology informed neural network_gaeta} are presented in Figures~\ref{gaeta_fig}(\textbf{a}) and~\ref{gWebb_fig}(\textbf{b}). For Italy and USA data, we used the following formulation for $\beta(t)$ in Algorithm~\ref{alg:Epidemiology informed neural network_gaeta} 
\begin{equation}
\beta(t) =  \begin{cases}
\beta_0 q_1 & 0 \leq t_j \leq 20\\
\beta_0 q_2 & 20 < t_j \leq 35\\
\beta_0 q_3 & 35 < t_j \leq 100\\
\beta_0 q_4 & 100 < t.
\end{cases}
\end{equation}

% new tables
\begin{table}[htbp]
\centering
\begin{tabular}{lcc}
\addlinespace
\toprule
{Parameters} & {Mean} & {Std} \\
\toprule
$\gamma$  & 0.00459 & 0.00013 \\
$\mu$  & 0.01202 & 0.00238 \\
$q_2$  & 0.29297 & 0.17053 \\
$q_3$  & 0.53369 & 0.16459 \\
$q_4$  & 0.49833 & 0.07585 \\
\midrule
\end{tabular}
\caption{Setting $q_1 = 1$, EINN Algorithm \ref{alg:Epidemiology informed neural network_gaeta} learns $q_2$, $q_3$, and $q_4$ for Italy data from 31 January 2020 to 11 December 2020}
\label{Table:pi1}
\end{table}

\begin{table}[htbp]
\centering
\begin{tabular}{lcc}
\addlinespace
\toprule
{Parameters} & {Mean} & {Std} \\
\toprule
$\gamma$  & 0.01338 & 0.00062 \\
$\mu$  & 0.01700 & 0.00704 \\
$q_2$  & 0.73672 & 0.19446 \\
$q_3$  & 0.84209 & 0.16659 \\
$q_4$  & 0.84166 & 0.09445 \\
\midrule
\end{tabular}
\caption{Setting $q_1 = 1$, EINN Algorithm \ref{alg:Epidemiology informed neural network_gaeta} learns $q_2$, $q_3$, and $q_4$ for USA data from 31 January 2020 to 11 December 2020}
\label{Table:pi2}
\end{table}

\section{Discussion}
\label{discuss}

\subsection{Mitigation Measures}
The COVID-19 infectious population surge witnessed in March and April 2020 around the world forced many countries to institute strict lockdown measures. 
This was largely successful in reducing the $\mathcal{R}_0$ in many countries, unfortunately, it also resulted in economic hardship, such that we seek other measures that also reduce the $\mathcal{R}_0$ to a number less than $1$. 
In recent months, the measures that are promoted in most countries include contact tracing, social distancing, and facial covering. 
The epidemiological meaning of each of the model parameters in Equation \eqref{Asymptomatic-SIRD_0} including $\xi$ are presented in Sections \ref{cparam}--\ref{ctrace}.

\subsection{Vaccination Efficacy}
In Figure~\ref{us_uk_vac}(\textbf{b}), using USA data, the mitigation effect of vaccination on the daily infectives is demonstrated. Implementing Algorithm~\ref{alg:Epidemiology informed neural networksird_cons}, we obtained $\kappa = 0.00184$, which is slightly different from the projection of $\kappa$ = $0.00305$, corresponding to $1$ million people vaccinated per day.
In Figure~\ref{us_uk_vac}(\textbf{a}), using United Kingdom data, we simulate the impact of vaccination on the daily reported infectives, using a smoothed daily vaccination data from 13 December 2020 to 5 February 2020 and smoothed daily reported infectives data.
We implement Algorithm~\ref{alg:Epidemiology informed neural networksird_cons} and we obtained $\kappa = 0.00305$. 
%Our model predicts that the daily infectives will be close to zero by April 2021.
We demonstrate the impact of increased social distancing together with the vaccination effort.
Social distancing corresponds to decreasing the transmission rate $\beta$.
Increased social distancing reduces the daily reported infectives but it extends the number of days daily infectives data is significant.

%In Figure~\ref{us_uk_vac}(\textbf{b}), using USA data, the mitigation effect of vaccination on the daily infectives is demonstrated.
%Implementing Algorithm \ref{alg:Epidemiology informed neural networksird_cons}, we obtained $\kappa = 0.00184$, which is slightly different from the projection of $\kappa$ = $0.00305$, corresponding to $1$ million people vaccinated per day.
%In Figure \ref{us_uk_vac}(\textbf{a}), using United Kingdom data, we simulate the impact of vaccination on the daily reported infectives, using a smoothed daily vaccination data from \hl{13 December 2020} to \hl{5 February 2020} and smoothed daily reported infectives data.
%We implement Algorithm \ref{alg:Epidemiology informed neural networksird_cons} and we obtained $\kappa = 0.00305$. 
%%Our model predicts that the daily infectives will be close to zero by April 2021.
%We demonstrate the impact of increased social distancing together with the vaccination effort.
%Social distancing corresponds to decreasing the transmission rate $\beta$.
%Increased social distancing reduces the daily reported infectives but it extends the number of days daily infectives data is significant.

\subsection{Time-Varying Transmission Rate}
%\noindent 
In Section \ref{gwebb_result}, the delayed-mitigation exponential time-varying transmission rate detects the impact of 2020 COVID-19 lockdown, as well as the other mitigation measures post-lockdown using the parameter $\eta$. 
It is however difficult to know if $\eta$ captures all the pattern in the time-varying transmission rate as demonstrated in \mbox{Figure~\ref{gWebb_fig}a,b}, i.e., whether or not Equation \eqref{webb_eqn} helps us to learn the most accurate form of $\beta$. 
For instance, the time-varying basic reproduction rate $\mathcal{R}_t$ is underestimated pre-lockdown in the USA data and overestimated pre-lockdown in Italy data.

%In Section \ref{gwebb_result}, the delayed-mitigation exponential time-varying transmission rate detects the impact of 2020 COVID-19 lockdown, as well as the other mitigation measures post-lockdown using the parameter $\eta$. 
%It is however difficult to know if $\eta$ captures all the pattern in the time-varying transmission rate as demonstrated in \mbox{Figure~\ref{gWebb_fig}a,b}, i.e., whether or not Equation \eqref{webb_eqn} helps us to learn the most accurate form of $\beta$. 
%For instance, the time-varying basic reproduction rate $\mathcal{R}_t$ is underestimated pre-lockdown in the USA data and overestimated pre-lockdown in Italy data. 
%$\mathcal{R}_t$ is overestimated post-lockdown. 

\subsection{Error Metrics for Data-Driven Simulation}
\begin{figure}[htbp]
\centering
\includegraphics[width=17.0cm]{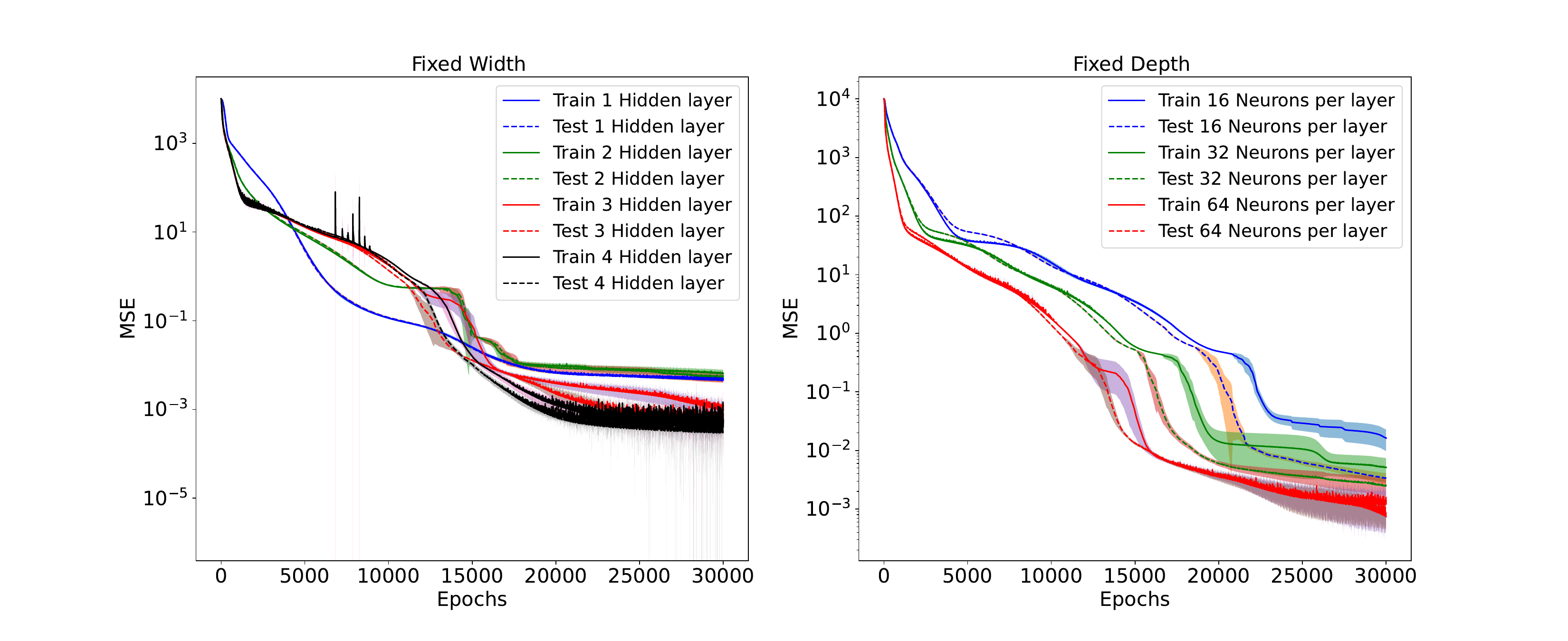}
\caption{Training and testing Errors in EINN for Italy data}
\label{Epoch_depth}
\end{figure}

The performance of the neural network training is demonstrated in Table~\ref{Table:er}, where the random and shuffle splits~\citep{Chollet2017} has been used to generate the training and testing dataset. The random split performed better than the shuffle split. In  Figure~\ref{Epoch_depth}, we present the training and testing MSE at different epochs, depths and widths. We observe that it is more beneficial to increase the width before increasing the depth~\citep{Jaehoon2017}.

\begin{table}[htbp]
\centering
\begin{tabular}{lcccc}
\addlinespace
\toprule
{\bf Data Split} & {\bf $R_2$ score} & {\bf MSE} & {\bf MAE} & {\bf Max Error}\\
\toprule
Random split  & $9.9994 \times 10^{-1}$ & $3.9365 \times 10^{-4}$ & $1.2440 \times 10^{-2}$ & $6.6720 \times 10^{-2}$  \\
 Shuffle split & $9.2104 \times 10^{-1}$ & $4.4006 \times 10^{-1}$ & $4.9789 \times 10^{-1}$ & $ 1.3683\times 10^{0}$  \\
\midrule
\end{tabular}
\caption{Error metrics for the infected cases $(I)$ using the random and shuffle splits for Italy COVID data, where we use 40\% of the dataset for testing.}
\label{Table:er}
\end{table}

\section{Conclusions}
\label{conclusion}
%%%%%%%%%%%%%%%%%%%%%%%%%%%%%%%%%%%%%%%%%%%%%%%%%%%%%%%%%%%%%%%%%%%%%%%%%%%%%

We have presented a data-driven deep-learning algorithm that discovers transmission rate patterns in an epidemiology model using cumulative and daily reported symptomatic infective and recovered data. 
The algorithm predicts asymptomatic infectives and asymptomatic recovered populations. 
The asymptomatic population is usually unreported in the publicly available data. 
We learn this population from symptomatic population data. 
It is demonstrated that a time-varying function models the nonlinear transmission rate.
The EINN algorithms presented, learns the nonlinear time-varying transmission rate without a pre-assumed pattern. 
This approach is useful when the dynamics of an epidemiological model is impacted by various mitigation measures. 
The algorithm can be adapted to most epidemiology models. 

In the proposed model, we have demonstrated the impact of public health actions on the transmission of COVID-19.
The effect of pharmaceutical mitigation measures such as vaccination is presented. 
Non-pharmaceutical mitigation measures such as early detection of symptomatic infectives population, contact tracing, and social distancing are promoted by showing their impact on the spread of COVID-19. 
This study is useful in the event of a pandemic such as COVID-19, where governmental interventions and public response and perceptions interfere in the interaction of the compartments in an epidemiology model.

\newpage

\appendix

%%%%%%%%%%%%%%%%%%%%%%%%%%%%%%%%%%%%%%%%%%%%%%%%%%%%%%%%%%%%%%%%%%%%%%%%%%%%%
\section{EINN Algorithm for Constant Transmission Rates} 
\label{cons_BGM}

We present the EINN Algorithm \ref{alg:Epidemiology informed neural networksird_cons} for the asymptomatic-SIR model with constant parameters.
That is, in Equation \eqref{Asymptomatic-SIRD_0}, we set $\beta(t)=\beta$.
The learned cumulative infectives and the recovered solution is matched against the cumulative infectives and recovered data.
In this algorithm, the parameters represent average rates.
We implement \mbox{Algorithm~\ref{alg:Epidemiology informed neural networksird_cons}} using publicly available \href{https://github.com/CSSEGISandData/COVID-19}{COVID-19 data} \cite{Dong2020}.

\begin{breakablealgorithm}
%\begin{algorithm}[H]
\caption{EINN algorithm for Asymptomatic-SIR model with constant parameters}
\label{alg:Epidemiology informed neural networksird_cons}
\begin{algorithmic}[1]

\State{Construct EINN}
 
{specify the input: $t_j$, $j=1,\ldots,M$}

{Initialize EINN parameter: $\theta$}

{Initialize the epidemiology and vaccination parameters: $\lambda=[\beta, \gamma, \mu, \xi, \kappa]$}

{Output layer: $S(t_j; \theta; \lambda)$, $I(t_j; \theta; \lambda)$, $J(t_j; \theta; \lambda)$, $R(t_j; \theta; \lambda)$, $U(t_j; \theta; \lambda)$, $j=1,\ldots,M$.}

\State{Specify the training set }

{Training data: using cubic spline, generate $\tilde{I}(t_j)$, $\tilde{R}(t_j)$, $j=1,\ldots,M$ and $\tilde{V}(t_j)$, $j=1,\ldots,M_{\kappa}$. from given dataset. } 

{Initialize the Asymptomatic population: $\tilde{J}(0) = (1-\xi) \tilde{I}(0)/\xi$ and $\tilde{U}(0) = (1-\xi) \tilde{R}(0)/\xi$.}

\State{Train the neural network}

{Specify an $MSE$ loss function:
\begin{equation}
 \begin{split}
MSE &= \frac{1}{M}\sum_{j=1}^{M}||I(t_j; \theta; \lambda) - \tilde{I}(t_j)||_2^2 + \frac{1}{M}\sum_{j=1}^{M}||R(t_j; \theta; \lambda) - \tilde{R}(t_j)||_2^2 \\
& +  \frac{1}{M_{\kappa}}\sum_{j=1}^{M_{\kappa}}||\kappa S(t_j;\theta;\lambda) - \tilde{V}(t_j)||_2^2\\
& + ||J(0;\theta;\lambda) - \tilde{J}(0)||_2^2 + ||U(0;\theta;\lambda) - \tilde{U}(0)||_2^2\\
& + \frac{1}{M}\sum_{i=1}^6\sum_{j=1}^{M}||L_i(t_j;\theta;\lambda)||_2^2.
 \end{split}
\end{equation} }

{Minimize the $MSE$ loss function:
compute $\argmin\limits_{\{\theta; \lambda\}}(MSE)$ using an optimizer such as the adam optimizer.}

\State{return EINN solution}

{$S(t_j;\theta;\lambda)$, $I(t_j;\theta;\lambda)$, $J(t_j;\theta;\lambda)$, $R(t_j;\theta;\lambda)$, $U(t_j;\theta;\lambda)$, $j=1,\ldots,M$.}

{parameters: $\beta, \gamma, \mu, \xi, \kappa$}. 

\end{algorithmic}
%\end{algorithm}
\end{breakablealgorithm}

\newpage

%%%%%%%%%%%%%%%%%%%%%%%%%%%%%%%%%%%%%%%%%%%%%%%%%%%%%%%%%%%%%%%%%%%%%%%%%%%%%
\section{EINN Algorithm for Time-Varying Transmission Rate} 
\label{A-SIR_ketch}

The time-varying transmission rate is non-constant in the presence of mitigation measures in the cumulative infective data.
In \cite{Liu2020, Liu20202}, it was shown that during the early phase of the COVID-19 pandemic when the cumulative infection population grew exponentially, the transmission rate was constant. 
This coincides with the period before any mitigation measure. Incorporating measures such as social distancing, lockdown and widespread adoption of facial covering in an epidemiology model is complex. 
We learn an exponentially decreasing transmission rate, we see that it takes the form of \mbox{Equation \eqref{webb_eqn}}. Our approach also detects various other post-lockdown mitigation measures.
We use EINN \mbox{Algorithm \ref{alg:Epidemiology informed neural network_gwebb}} to learn $\beta(t)$.

\begin{breakablealgorithm}
%\begin{algorithm}[H]
%\footnotesize
\caption{EINN algorithm for Asymptomatic-SIR model with delayed-mitigation exponential time-varying transmission rate}
\label{alg:Epidemiology informed neural network_gwebb}
\begin{algorithmic}[1]

\State{Construct EINN}
 
{specify the input: $t_j$, $j=1,\ldots,M$}

{Initialize EINN parameter: $\theta$}

{Initialize the epidemiology and vaccination parameters: $\lambda=[\gamma, \mu, \kappa]$}

{Output layer: $S(t_j; \theta; \lambda)$, $I(t_j; \theta; \lambda)$, $J(t_j; \theta; \lambda)$, $R(t_j; \theta; \lambda)$, $U(t_j; \theta; \lambda)$, $j=1,\ldots,M$}

\State{construct neural network: $\beta$}

{specify the input: $t_j$, $j=1,\ldots,M$}

{Initialize the neural network parameter: $\phi$}

{Specify $\beta_0$ obtained by nonlinear regression of early cumulative infective population data}

{Initialize the exponential decay parameter: $\eta$}

{Output layer: $\beta(t_j;\phi;\eta)$  
\begin{equation}
\beta(t_j;\phi;\eta) =  \begin{cases}
\beta_0 & 0 \leq t_j \leq M_{\beta}\\
\beta_0  \exp{(-\eta\beta(t_j;\phi;\eta))} & M_{\beta} < t_j,
\end{cases}
\end{equation} }

%\State{specify the Output: $S(t_j; \theta; \lambda)$, $I(t_j; \theta; \lambda)$, $J(t_j; \theta; \lambda)$, $R(t_j; \theta; \lambda)$, $U(t_j; \theta; \lambda)$, $j=1,\ldots,M_{\kappa}$, $\beta(t_j;\phi;\eta)$.}

\State{Specify EINN training set }

{Training data: using cubic spline, generate $\tilde{I}(t_j)$ and $\tilde{R}(t_j)$, $j=1,\ldots,M$.}

{Set $\xi$ to the value obtained for $\xi$ from EINN Algorithm \ref{alg:Epidemiology informed neural networksird_cons} }

{Initialize the Asymptomatic population: $\tilde{J}(0) = (1-\xi) \tilde{I}(0)/\xi$ and $\tilde{U}(0) = (1-\xi) \tilde{R}(0)/\xi$.}

\State{Train the neural networks}

{Specify an $MSE$ loss function:
\begin{equation}\label{loss_func_beta_2}
 \begin{split}
MSE &= \frac{1}{M}\sum_{j=1}^{M}||I(t_j; \theta;\lambda) - \tilde{I}(t_j)||_2^2 + \frac{1}{M}\sum_{j=1}^{M}||R(t_j; \theta;\lambda) - \tilde{R}(t_j)||_2^2 \\
& +  \frac{1}{M_{\beta}}\sum_{j=1}^{M_{\beta}}||\beta(t_j;\phi;\eta) - \beta_0  ||_2^2\\
& +  \frac{1}{M_{\kappa}}\sum_{j=1}^{M_{\kappa}}||\kappa S(t_j;\theta) - \tilde{V}(t_j)||_2^2\\
& + ||J(0;\theta;\lambda) - \tilde{J}(0)||_2^2 + ||U(0;\theta;\lambda) - \tilde{U}(0)||_2^2\\
& + \frac{1}{M}\sum_{i=1}^6\sum_{j=1}^{M}||L_i(t_j;\theta;\phi;\lambda;\eta)||_2^2.
 \end{split}
\end{equation} }

{Minimize the $MSE$ loss function:
compute $\argmin\limits_{\{\theta; \phi; \lambda; \eta\}}(MSE)$ using an optimizer such as the adam optimizer.}

\State {return EINN solution}

{$S(t_j;\theta;\lambda)$, $I(t_j;\theta;\lambda)$, $J(t_j;\theta;\lambda)$, $R(t_j;\theta;\lambda)$, $U(t_j;\theta;\lambda)$, $j=1,\ldots,M$.}

{epidemiology parameters: $\gamma$ and $\mu$}

{vaccination parameter: $\kappa$ }

\State {return time-varying epidemiology parameter:}

{$\beta(t_j;\phi;\eta)$, $j=1,\ldots,M$. }

{Rate of human contact decrease: $\eta$. }

\end{algorithmic}
%\end{algorithm}
\end{breakablealgorithm}

\newpage

\section{EINN Algorithm for Piecewise time-varying transmission rate}
A piecewise time-varying transmission rate \eqref{gaeta_eqn} is used to learn a time-dependent transmission rate $\beta$ in eq. \eqref{Asymptomatic-SIRD_0}. 

\begin{breakablealgorithm}
%\begin{algorithm}[H]
%\scriptsize
\caption{EINN algorithm for Asymptomatic-SIR model with piecewise time-varying transmission rate}
\label{alg:Epidemiology informed neural network_gaeta}
\begin{algorithmic}[1]

\State{Construct EINN}
 
{specify the input: $t_j$, $j=1,\ldots,M$}

{Initialize EINN parameter: $\theta$}

{Initialize the epidemiology and vaccination parameters: $\lambda=[\gamma, \mu, \kappa]$}

{Output layer: $S(t_j; \theta; \lambda)$, $I(t_j; \theta; \lambda)$, $J(t_j; \theta; \lambda)$, $R(t_j; \theta; \lambda)$, $U(t_j; \theta; \lambda)$, $j=1,\ldots,M$}

\State{construct neural network: $\beta$}

{specify the input: $t_j$, $j=1,\ldots,M$}

{Initialize the neural network parameter: $\phi$}

{Specify $\beta_0$ obtained by nonlinear regression of early cumulative infective population data}

{Initialize the decay parameters: $q_1,q_2,q_3,q_4, \ldots, q_n$}

{Output layer: $\beta(t_j;\phi;q_1,q_2,q_3,q_4, \ldots, q_n)$  
\begin{equation}
\beta(t_j;\phi;q_1,q_2,q_3,q_4, \ldots, q_n) =  \begin{cases}
\beta_0 q_1 \beta(t_j;\phi;q_1)& 0 \leq t_j \leq M_{1}\\
\beta_0 q_2 \beta(t_j;\phi;q_2)& M_{1} < t_j \leq M_{2}\\
\beta_0 q_3 \beta(t_j;\phi;q_3)& M_{2} < t_j \leq M_{3}\\
\beta_0 q_4 \beta(t_j;\phi;q_4)& M_{3} < t_j \leq M_{4}\\
& \vdots\\
\beta_0 q_n \beta(t_j;\phi;q_n)& M_{n} < t_j,
\end{cases}
\end{equation} }

%\State{specify the Output: $S(t_j; \theta; \lambda)$, $I(t_j; \theta; \lambda)$, $J(t_j; \theta; \lambda)$, $R(t_j; \theta; \lambda)$, $U(t_j; \theta; \lambda)$, $j=1,\ldots,M_{\kappa}$, $\beta(t_j;\phi;\eta)$.}

\State{Specify EINN training set }

{Training data: using cubic spline, generate $\tilde{I}(t_j)$ and $\tilde{R}(t_j)$, $j=1,\ldots,M$.}

{Set $\xi$ to the value obtained for $\xi$ from EINN Algorithm \ref{alg:Epidemiology informed neural networksird_cons} }

{Initialize the Asymptomatic population: $\tilde{J}(0) = (1-\xi) \tilde{I}(0)/\xi$ and $\tilde{U}(0) = (1-\xi) \tilde{R}(0)/\xi$.}

\State{Train the neural networks}

{Specify an $MSE$ loss function:
\begin{equation}\label{loss_func_beta_3}
 \begin{split}
MSE &= \frac{1}{M}\sum_{j=1}^{M}||I(t_j; \theta;\lambda) - \tilde{I}(t_j)||_2^2 + \frac{1}{M}\sum_{j=1}^{M}||R(t_j; \theta;\lambda) - \tilde{R}(t_j)||_2^2 \\
& +  \sum_{i=1}^{n}\frac{1}{M_{i}}\sum_{j=1}^{M_{i}}||\beta(t_j;\phi;\eta) - \beta_0  ||_2^2\\
& +  \frac{1}{M_{\kappa}}\sum_{j=1}^{M_{\kappa}}||\kappa S(t_j;\theta) - \tilde{V}(t_j)||_2^2\\
& + ||J(0;\theta;\lambda) - \tilde{J}(0)||_2^2 + ||U(0;\theta;\lambda) - \tilde{U}(0)||_2^2\\
& + \frac{1}{M}\sum_{i=1}^6\sum_{j=1}^{M}||L_i(t_j;\theta;\phi;\lambda;\eta)||_2^2.
 \end{split}
\end{equation} }

{Minimize the $MSE$ loss function:
compute $\argmin\limits_{\{\theta; \phi; \lambda; \eta\}}(MSE)$ using an optimizer such as the adam optimizer.}

\State {return EINN solution}

{$S(t_j;\theta;\lambda)$, $I(t_j;\theta;\lambda)$, $J(t_j;\theta;\lambda)$, $R(t_j;\theta;\lambda)$, $U(t_j;\theta;\lambda)$, $j=1,\ldots,M$.}

{epidemiology parameters: $\gamma$ and $\mu$}

{vaccination parameter: $\kappa$ }

\State {return time-varying epidemiology parameter:}

{$\beta(t_j;\phi;q_1,q_2,q_3,q_4, \ldots, q_n)$, $j=1,\ldots,M$. }

{Rate of human contact decrease: $q_1,q_2,q_3,q_4, \ldots, q_n$. }

\end{algorithmic}
%\end{algorithm}
\end{breakablealgorithm}

%\section*{References}

\bibliography{mybibfile}

\end{document}